\documentclass[12pt,a4paper]{article}
\usepackage[truedimen,margin=30mm]{geometry} 

\usepackage{mathrsfs}
\usepackage{amssymb}
\usepackage{amsmath}
\usepackage{ascmac}
\usepackage{amsthm}
\usepackage[dvips]{graphicx}
\usepackage{natbib}
\usepackage{setspace}
\usepackage{times}
\usepackage{comment}
\usepackage{url}

\usepackage{color}
\usepackage{float}

\usepackage{titlesec}
\titleformat*{\section}{\large\bfseries}
\titleformat*{\subsection}{\it}

\newtheorem{df}{Definition}
\newtheorem{thm}{Theorem}
\newtheorem{lem}{Lemma}

\newtheorem{as}{Assumption}
\def\bD{{\boldsymbol{D}}}
\def\bd{{\boldsymbol{d}}}

\title{{\bf Difference-in-Differences under Local Dependence on Networks}\footnote{\today}}

\date{}

\begin{document}

\maketitle
\doublespacing

\vspace{-1.5cm}
\begin{center}
{\large Akihiro Sato$^1$ and Shonosuke Sugasawa$^2$}

\medskip

\medskip
\noindent
$^1$Graduate School of Economics, Keio University\\
$^2$Faculty of Economics, Keio University\\
\end{center}

\vspace{0.3cm}
\begin{center}
{\bf \large Abstract}
\end{center}

\vspace{-0cm}
Estimating causal effects under interference, where the stable unit treatment value assumption is violated, is critical in fields such as regional and public economics. Much of the existing research on causal inference under interference relies on a pre-specified ``exposure mapping.'' This paper focuses on difference-in-difference and proposes a nonparametric identification strategy for direct and indirect average treatment effects under local interference on an observed network. In particular, we proposed a new concept of an indirect effect measuring the total outward influence of the intervension. Based on parallel trends assumption conditional on the neighborhood treatment vector, we develop inverse probability weighted and doubly robust estimators. We establish their asymptotic properties, including consistency under misspecification of nuisance models under some regularity conditions. Simulation studies and an empirical application demonstrate the effectiveness of the proposed method. 


\newpage
\section{Introduction}

Estimating the causal effect of an intervention remains a central challenge in empirical research. While the Difference-in-Differences (DID) framework has long served as a cornerstone for this purpose, its foundational assumptions have recently received significant attention. Recent literature has demonstrated that in settings with staggered treatment adoption, conventional DID estimators can be severely biased due to the contamination of control groups by previously treated units (e.g., \citealp{Goodman-Bacon2021}; \citealp{Callaway2021}; \citealp{Sun2021}; \citealp{deChaisemartin2020}). A parallel but distinct source of control group contamination is spillover from contemporaneously treated units on a network. In many applications, such as the evaluation of place-based policies or marketing campaigns, treatment effects are not confined to the treated units but spill over to others. This phenomenon, known as ``interference" \citep{Cox1958}, constitutes a violation of the stable unit treatment value assumption (SUTVA; \citealp{Rubin1980}), a fundamental premise in causal inference.

When interference is present, the control group fails to serve as a valid counterfactual for the absence of treatment. Consequently, a simple comparison between treated and control groups yields a biased estimate of the true causal effect \citep[e.g.,][]{perez2014assessing,Butts2023}. This bias can be severe, potentially reversing the sign of the effect and leading to fundamentally flawed policy implications \citep{leung2024identifying}.
However, causal inference under interference is inherently challenging because one must account for $2^N$ possible treatment assignments for $N$ units. This complexity makes it nearly impossible to identify meaningful causal effects without imposing structural assumptions \citep{Manski2013}. Early influential work addressing this issue includes studies on social interactions (e.g., \citealp{Manski1993}) and vaccine efficacy (e.g., \citealp{Hudgens2008}).

A more flexible foundation is the concept that a unit's outcome may depend on the entire vector of treatment assignments, formalized by \citet{Manski2013} as the effective treatment. \citet{Aronow2017} introduced a systematic framework known as exposure mapping, which summarizes the high-dimensional treatment vector into a low-dimensional ``exposure level." Researchers can define interference and estimate effects conditional on exposure levels (e.g., \citealp{Ugander2013}; \citealp{Basse2018}). A major practical limitation is the need to pre-specify the interference structure (e.g., the number of treated neighbors required to trigger an effect), which must be known and correctly specified a priori. Still limited, however, is a flexible DID framework that achieves identification under interference without pre-specifying this structure.
Social and economic interference is often complex and nonlinear, making it rare for researchers to know the correct functional form in advance. The reliance on a known exposure mapping poses a significant risk; in complex real-world networks, misspecification of this mapping can lead to severe bias in causal effect estimates. The difficulty of correctly specifying this mapping and the consequences of misspecification have been highlighted in recent theoretical work, such as \citet{Savje2024}, which investigates the conditions for identification under misspecification and underscores the severity of this issue.
While recent advances in network causal inference, such as the auto-doubly robust framework proposed by \citet{liu2025auto}, have successfully addressed general interference using Markov random fields and autoregressive models, these methods often rely on specific parametric structures for nuisance functions. In contrast, our approach leverages the temporal dimension of panel data to identify causal effects nonparametrically, thereby avoiding structural assumptions on the spillover mechanism and providing a more robust alternative in settings where it is poorly understood (e.g., \citealp{Xu2025}).

This paper proposes a novel identification strategy that exploits the panel data structure to achieve nonparametric identification without imposing a known exposure mapping. We condition on the neighborhood treatment vector under a conditional parallel trends assumption; the approach yields identification without functional form assumptions on the spillover structure. Estimation relies on flexible methods (IPW and doubly robust) that remain feasible when conditioning on this vector.

Recent work integrates interference into the DID framework. Early studies in spatial econometrics proposed spatial DID models that account for spillovers from geographically proximate units \citep{Delgado2015}. More closely related to our work is a recent line of research extending this integration to general network structures. While studies such as \citet{Xu2025} and \citet{Butts2023} make important contributions, they typically rely on a pre-specified exposure mapping. Our research aligns with the recent trend of relaxing strong assumptions regarding the functional form of interference (e.g., \citealp{Leung2022}). We contribute to this growing literature by embedding a nonparametric identification strategy—conditioning on the neighborhood treatment vector itself—within the DID framework, which remains the most widely used method for policy evaluation in observational settings.

A distinct contribution is our definition of the indirect effect. Much of the literature, including \citet{Xu2025}, defines the indirect (or spillover) effect as the difference in a unit's own outcome from varying exposure levels—\textit{inward} spillover: ``How would my outcome have changed if my neighborhood's treatment status had been different?" (see also \citealp{TchetgenTchetgen2012}). We instead define the average indirect treatment effect on the treated (AITT) as the effect of treating a given unit on its neighbors' outcomes, i.e., \textit{outward} spillover: ``By intervening on unit $i$, how much did the outcomes of its neighbors $j$ change on average?" The AITT directly quantifies policy externalities and is distinct from mediated effects (e.g., \citealp{Sobel2006}).

We propose IPW (and doubly robust) estimators that implement this identification and establish their asymptotic theory. For inference we adopt the $\psi$-dependence framework of \citet{Kojevnikov2021}, which delivers asymptotic normality when observations are dependent on a single large network, balancing network density and dependence decay; standard i.i.d.\ asymptotics do not apply under network dependence.
Conditioning on the neighborhood treatment vector can reduce efficiency (curse of dimensionality) relative to correctly specified exposure mapping, especially in dense networks or small samples. Our method is thus complementary: when the interference structure is known, exposure mapping is preferred for efficiency; when it is unknown, our nonparametric identification with flexible parametric estimation offers a robust alternative.

This paper is organized as follows. 
Section 2 presents the identification framework. Section 3 proposes estimators and inference methods. Section 4 establishes the asymptotic properties of the estimators. Section 5 reports simulation evidence, Section 6 presents an empirical application, and Section 7 gives concluding remarks.
The Python code demonstrating the proposed method is available at GitHub repository (\url{https://github.com/S10aki95/DIDUnderLocalDependenceOnNetworks}).

\section{Identification}

We define the setup, potential outcomes, and causal parameters (ADTT and AITT), state the identifying assumptions—including an extension of parallel trends to the network—and establish the main identification results.

\subsection{Setup and potential outcomes}

We consider a population of $n$ units, $i\in I_n$, connected by a network with adjacency matrix $A$, where $l_A(i,j)$ is the shortest path length. The data are two-period panel ($t=1,2$) with treatment indicator $D_i$ and outcome $Y_{ti}$; inference is conditional on fixed attributes $z_i$ (finite population perspective). 
We consider a setting characterized by interference (spillover effects), violating the stable unit treatment value assumption (SUTVA). We specifically assume that interference is local and occurs along the network structure.
To fix ideas, we define the neighborhood of unit $i$, denoted by $N_i$, as the set of its $L$ nearest neighbors on the network, where $L$ is a fixed number chosen by the researcher; this keeps the conditioning set dimension fixed for estimation. Specifically, for each unit $i$, we sort all units $j \in I_n \setminus \{i\}$ by their network distance $l_A(i,j)$ in ascending order, and take the first $L$ units to form the neighborhood $N_i$. Let $\boldsymbol{D}_{N_i} \in \{0,1\}^{L}$ denote the treatment vector of these neighbors, sorted by their distance from unit $i$. That is, for $k = 1, \ldots, L$, the $k$-th component of $\boldsymbol{D}_{N_i}$ corresponds to the treatment indicator of the $k$-th nearest neighbor in $N_i$.

For the asymptotic theory developed in Section 4, we assume that interference is confined to a distance of $K$, where $K$ denotes the theoretical interference range (the maximum network distance within which units can affect each other). We define the neighborhood within distance $K$ for the $i$th unit as
$$
N(i;K) = \{ j\in I_n\setminus\{i\} \mid l_A(i,j)\leq K\}.
$$
For any unit $j$, we also define $N(j;K)^+ = N(j;K) \cup \{j\}$ as the extended neighborhood that includes unit $j$ itself. We also define $N^{\partial}(i; s) = \{ j \in I_n \mid l_A(i, j) = s \}$ as the set of units at exactly distance $s$ from unit $i$. When considering the treatment vector for unit $j$'s neighborhood excluding unit $i$, we denote this as $\boldsymbol{d}_{N(j;K)}^{+, -i} \in \{0, 1\}^{|N(j;K)^+ \setminus \{i\}|}$, which represents the treatment assignment vector for all units in $j$'s extended neighborhood except for unit $i$.

Under this setting, we define $y_{ti}(d_i, \bd_{N_i})$ as the potential outcome for unit $i$ when its own treatment status is fixed at $d_i\in \{0,1\}$ and its neighborhood treatment vector is fixed at $\bd_{N_i}\in \{0,1\}^{L}$.
This formulation is general and encompasses structures such as $y_{2i}(d_i, \bd_{N_i}) = m(Y_{1i}, d_i, \bd_{N_i}, \epsilon_i)$, where $\epsilon_i$ is an error term. A key feature of this formulation is that it allows the entire combination of neighborhood treatments, $\bd_{N_i}$, to affect the outcome, without requiring a specific functional form like an exposure mapping, which is often assumed in literature \citep{Hudgens2008,Ugander2013}. A fixed-length vector $\bd_{N_i}$ based on the $L$ nearest neighbors keeps the conditioning set dimension constant across units.

\subsection{Causal estimands}

Under the setup described in the previous section, we first define the direct effect.
\begin{df}
The average direct treatment effect on the treated (ADTT), given by
$$
\tau_{\mathrm{dir}} = \frac{1}{n} \sum_{i=1}^n E[y_{2i}(1, \bD_{N_i}) - y_{2i}(0, \bD_{N_i}) \mid D_i=1, z_i],
$$
captures the expected direct effect of treatment for the treated subpopulation, marginalized over the observed neighborhood treatment vectors. The expectation $E[\cdot]$ is over stochastic unobservables for unit $i$, conditional on treatment status and fixed attributes $z_i$.
\end{df}
This definition ensures that the estimand is well-defined under interference; therefore $\tau_{\mathrm{dir}}$ is interpreted as the expected outcome difference for treated units between $D_i=1$ and $D_i=0$, holding $\bD_{N_i}$ fixed. The formulation is standard in the literature \citep[e.g.,][]{Aronow2017, Xu2025}. 
While the definition of the estimand is the same, the identification and estimation strategies differ significantly. Prior work often relies on an exposure mapping that reduces the high-dimensional treatment assignment vector to a low-dimensional exposure level \citep[e.g.][]{Hudgens2008,Aronow2017,Xu2025}. Our key contribution regarding ADTT is that we achieve nonparametric identification without assuming a known exposure mapping. We do this by conditioning directly on the observed neighborhood treatment vector $\bD_{N_i}$. This avoids the risk of misspecification inherent in exposure mapping and allows nonparametric identification of the direct effect.

For the indirect effect we introduce the Average Indirect Treatment Effect on the Treated (AITT), which differs from definitions commonly used in prior studies.

\begin{df}
The Average Indirect Treatment Effect on the Treated (AITT), given by
$$
\tau_{\mathrm{ind}} = \frac{1}{n} \sum_{i=1}^n \left( \frac{1}{|N_i|} \sum_{j \in N_i} E\Big[\Delta y_{2j} | D_i = 1, z_i, z_j \Big] \right),
$$
quantifies the average externality on neighbors from treating unit $i$, where $\Delta y_{2j} = y_{2j}(D_j, \bD_{N_j}^{(1)}) - y_{2j}(D_j, \bD_{N_j}^{(0)})$, and $\bD_{N_j}^{(d)}$ is the treatment vector of $j$'s neighbors with $D_i=d\in \{0,1\}$. The expectation $E[\cdot]$ is over stochastic unobservables of neighbor $j$, conditional on $z_i$ and $z_j$.
\end{df}

The inner summation captures the effect on each neighbor; the outer aggregation is the ``spillover footprint'' of treating unit $i$—the total outward influence of the intervention.
This definition is different from the standard definitions of indirect (or spillover) effects found in the literature \citep[e.g.,][]{TchetgenTchetgen2012, Xu2025}. Existing definitions typically focus on the effect of changes in neighborhood exposure on a unit's \textit{own} outcome (i.e., ``How would my outcome change if my neighbors' treatment status were different?"). 
Our AITT instead captures the outgoing spillover effect generated by the treatment of unit $i$.

\subsection{Identification results}

To identify these parameters, we require an assumption about the unobserved counterfactuals. We extend the standard parallel trends assumption in DID to accommodate network interference. 

\begin{as}[No anticipation] \ 
\label{as:NoAnticipation}
$E[y_{1i} | \bD_{I_n}, z_i] = y_{1i} (z_i)$.
\end{as}

\begin{as}[Conditional parallel trends under interference]
\label{as:trend}\
For all $i \in I_n$ and $\bd_{N_i} \in \{0,1\}^{L}$, it holds that 
\begin{align*}
& E[y_{2i}(0, \bd_{N_i}) - y_{1i}  | D_i=1, \bD_{N_i} = \bd_{N_i}, z_i] \\
& = E[y_{2i}(0, \bd_{N_i}) - y_{1i}  | D_i=0, \bD_{N_i} = \bd_{N_i}, z_i]
\end{align*}
Furthermore, for all $j \in N_i$, and $ \boldsymbol{d}_{N_j}^{-i} \in \{0, 1\}^{L-1} $, it holds that  
\begin{align*}
& E[y_{2j}(D_i=0, D_j, \boldsymbol{D}_{N_j}^{-i} = \boldsymbol{d}_{N_j}^{-i}) - y_{1j} \mid D_i=1, D_j, \boldsymbol{D}_{N_j}^{-i} = \boldsymbol{d}_{N_j}^{-i}, z_i, z_j] \\
& = E[y_{2j}(D_i=0, D_j, \boldsymbol{D}_{N_j}^{-i} = \boldsymbol{d}_{N_j}^{-i}) - y_{1j} \mid D_i=0, D_j, \boldsymbol{D}_{N_j}^{-i} = \boldsymbol{d}_{N_j}^{-i}, z_i, z_j]
\end{align*}
\end{as}

\begin{as}[Sufficiency and locality of L-neighborhood]
\label{as:neighborhood_sufficiency}\
We assume that the chosen number of neighbors $L$ is sufficient to capture the relevant interference. Furthermore, to ensure the validity of the asymptotic properties derived based on network distance, we assume that for all $i \in I_n$ and for all $j \in N_i$, the network distance satisfies $l_A(i, j) \leq K$. 
That is, the $L$-neighborhood $N_i$ is contained within the theoretical interference range $K$: $N_i \subseteq \{j \in I_n \setminus \{i\} \mid l_A(i, j) \leq K\}$.
\end{as}

Assumption~\ref{as:NoAnticipation} is known as ``no-anticipation assumption" that requires that the potential outcome in the first time period prior to treatment does not depend on treatments.
Assumption~\ref{as:trend} states that, conditional on the treatment status of all other units in the relevant neighborhood, the expected outcome trend under no treatment ($D_i=0$) is the same regardless of the actual treatment status of unit $i$. 
This is a natural extension of the standard conditional parallel trends assumption often employed in DID analysis with covariates \citep{abadie2005semiparametric}. In the standard conditional parallel trends assumption, parallel trends are assumed to hold conditional on a set of observed covariates. Here, we extend this logic to the network setting by conditioning on the neighborhood treatment vector $\bD_{N_i}$. By controlling for the exact configuration of neighborhood treatments, we substitute a structural assumption on the interference function with an assumption on the counterfactual trends, thereby enabling identification without an exposure mapping.
While this assumption is stronger than the standard unconditional parallel trends assumption, it is analogous to assumptions made in other recent work on nonparametric identification under network interference. For instance, \citet{Leung2022} employs a similar conditioning strategy (conditioning on neighborhood covariates and treatment assignments) to achieve identification in a cross-sectional setting.

Assumption~\ref{as:neighborhood_sufficiency} bridges the gap between the theoretical interference range $K$ and the practical implementation using a fixed number of neighbors $L$. It ensures that the chosen neighborhood $N_i$ captures all relevant interference effects (sufficiency) and that the asymptotic theory based on network distance remains valid (locality). Specifically, by requiring that all units in the $L$-neighborhood $N_i$ are within distance $K$ from unit $i$, this assumption guarantees that the estimators based on $L$-neighborhoods satisfy the mixing conditions and weak dependence properties derived under the theoretical interference range $K$. This assumption is reasonable in practice when $L$ is chosen to be sufficiently large relative to the network structure and the expected interference range.

Under Assumptions~\ref{as:NoAnticipation}, \ref{as:trend}, and \ref{as:neighborhood_sufficiency}, both ADTT and AITT are identified as in the following theorems. Proofs are in the Appendix. 

\begin{thm}
\label{thm:identification_ADTT}
Under Assumptions~\ref{as:NoAnticipation}, \ref{as:trend}, and \ref{as:neighborhood_sufficiency}, ADTT is identified as
$$
\tau_{\mathrm{dir}} = \frac{1}{n} \sum_{i=1}^n E \left[ \frac{D_i - e_i(\bD_{N_i}, z_i)}{\pi_i(z_i) \{1-e_i(\bD_{N_i}, z_i)\}} ( Y_{2i} - Y_{1i} ) \mid z_i \right],
$$
where $\pi_i(z_i) \equiv P(D_i=1|z_i)$ is the marginal probability of $D_i=1$ conditional on $z_i$, and $e_i(\bD_{N_i}, z_i) = P(D_i = 1 | \bD_{N_i}, z_i)$ is the propensity score conditional on the neighborhood treatment vector $\bD_{N_i}$ and $z_i$. Note that $e_i$ is a function of both $\bD_{N_i}$ and $z_i$. This weighting scheme adjusts not only for individual selection into treatment but also for the specific configuration of the neighborhood treatments, ensuring comparisons between units with similar exposure to local interference. The weight $e_i/\pi_i$ marginalizes over the distribution of $\bD_{N_i}$ given $D_i=1$, ensuring that the estimand captures the average direct effect on the treated subpopulation as defined in Definition 1.
\end{thm}

\begin{thm}
\label{thm:identification_AITT}
Under Assumptions~\ref{as:NoAnticipation}, \ref{as:trend}, and \ref{as:neighborhood_sufficiency}, AITT is identified as
$$
\tau_{\mathrm{ind}} = \frac{1}{n} \sum_{i=1}^n \left( \frac{1}{|N_i|} \sum_{j \in N_i} E \left[ \frac{D_i - e'_{ij}(D_j, \bD_{N_j}^{-i}, z_i, z_j)}{\pi_i(z_i)\{1-e'_{ij}(D_j, \bD_{N_j}^{-i}, z_i, z_j)\}} (Y_{2j} - Y_{1j}) \mid z_i, z_j \right] \right),
$$
where $e'_{ij}(D_j, \bD_{N_j}^{-i}, z_i, z_j) = P(D_i = 1 | D_j, \bD_{N_j}^{-i}, z_i, z_j)$ is the propensity of unit $i$ being treated, conditional on the treatment status of its neighbor $j$, $j$'s other neighbors, and the fixed attributes $z_i$ and $z_j$. Note that $e'_{ij}$ is a function of $D_j$, $\bD_{N_j}^{-i}$, $z_i$, and $z_j$.
\end{thm}

\section{Estimation and Inference}

The identification results in Section 2 lead directly to IPW (and doubly robust) estimators for ADTT and AITT; this section details their implementation and inference.
The identification results in Section 2 are nonparametric in that they require no functional form assumption on the interference structure and condition only on the neighborhood treatment vector $\bD_{N_i}$ (and covariates). In finite samples, however, conditioning on a high-dimensional vector entails the curse of dimensionality. To maintain practical feasibility and statistical performance, we therefore implement estimation using flexible parametric models—specifically, logistic regression for propensity scores and linear regression for outcome regressions—that use the full vector of neighbor treatments (e.g., distance-ranked indicators $D_{(k)}$) as covariates. These specifications can be viewed as parametric approximations to the nonparametric conditional expectations; researchers may alternatively employ machine learning methods for the nuisance functions when appropriate. The key point is that the \textit{identification} strategy remains nonparametric, whereas the \textit{estimation} step uses flexible parametric (or ML) approximations to make the approach practicable.

\subsection{Inverse probability weighted estimator}
Theorems \ref{thm:identification_ADTT} and \ref{thm:identification_AITT} yield the following inverse probability weighted (IPW) estimators for ADTT and AITT: 
\begin{equation}\label{eq:IPW}
\widehat{\tau}_{\mathrm{dir}} = \frac{1}{n} \sum_{i=1}^n \frac{D_i - \widehat{e}_i(\bD_{N_i}, z_i)}{\widehat{\pi}_i(z_i)\{1 - \widehat{e}_i(\bD_{N_i}, z_i)\}} (Y_{2i} - Y_{1i}),
\end{equation}
\begin{equation}
\widehat{\tau}_{\mathrm{ind}} = \frac{1}{n} \sum_{i=1}^n \left\{ \frac{1}{|N_i|} \sum_{j \in N_i} \left\{ \frac{D_i - \widehat{e}'_{ij}(D_j, \bD_{N_j}^{-i}, z_i, z_j)}{\widehat{\pi}_i(z_i)\{1 - \widehat{e}'_{ij}(D_j, \bD_{N_j}^{-i}, z_i, z_j)\}} (Y_{2j} - Y_{1j}) \right\} \right\}.
\end{equation}
Here, $\widehat{\pi}_i(z_i)$, $\widehat{e}_i(\bD_{N_i}, z_i)$, and $\widehat{e}'_{ij}(D_j, \bD_{N_j}^{-i}, z_i, z_j)$ are estimators of $\pi_i(z_i)$, $e_i(\bD_{N_i}, z_i)$ and $e'_{ij}(D_j, \bD_{N_j}^{-i}, z_i, z_j)$,  respectively. 

We estimate propensity scores using logistic regression. For ADTT, the model conditions on unit $i$'s covariates $z_i$ and the treatment statuses of its $L$ nearest neighbors, $\bD_{N_i}$, sorted by distance:
$$
\text{logit}(P(D_i=1)) = \alpha + \beta^\top z_i + \sum_{k=1}^L \gamma_k D_{(k)}.
$$
This specification is a flexible parametric approximation to $P(D_i=1\mid \bD_{N_i}, z_i)$ that avoids pre-specifying an exposure mapping but remains feasible in finite samples. By including the treatment indicators $D_{(k)}$ individually rather than aggregating them (e.g., into a sum), the model naturally captures heterogeneous spillover effects where the impact decays with distance or depends on the specific proximity of treated neighbors.
Similarly, for AITT, we estimate $e'_{ij}$ conditioning on covariates $z_i,z_j$, neighbor's treatment $D_j$, and the distance-sorted treatments of $j$'s other neighbors $\bD_{N_j}^{-i}$.
In both cases, the feature vectors have a fixed dimensionality determined by the number of covariates in $z_i$ (and $z_j$ for AITT) plus $L$ neighbor treatment indicators, ensuring computational feasibility while maintaining the nonparametric \textit{identification} strategy (with parametric models used for \textit{estimation}). The specific covariates used in the real data application are detailed in Section 6. 

Our proposed ADTT estimator differs from existing IPW estimators under interference, such as the one proposed by \citet{Xu2025}. While both utilize an IPW structure, the key difference lies in the conditioning set for the propensity score. Existing methods typically estimate the propensity score conditional on a low-dimensional, pre-specified exposure mapping $G_i$, i.e., $P(D_i=1|G_i)$. In contrast, our estimator uses the propensity score conditional on the full neighborhood treatment vector, $e_i = P(D_i = 1 | \bD_{N_i})$. This approach, which follows our nonparametric identification strategy, avoids the risk of misspecifying the exposure mapping, leading to more robust estimation of the direct effect.
However, this robustness comes at a cost. Estimating the propensity score conditional on a high-dimensional vector $\bD_{N_i}$ can be challenging, particularly when the number of neighbors $L$ is large (the curse of dimensionality). This may lead to reduced statistical efficiency compared to methods that correctly specify a low-dimensional exposure mapping. If researchers have strong prior knowledge about the functional form of interference, using an exposure mapping approach might be more efficient. When such knowledge is unavailable, our estimators that implement the nonparametric identification offer a robust alternative.

\subsection{Doubly robust estimator}

While the IPW estimators introduced above are based on our nonparametric identification strategy, they depend on the correct specification of the propensity score models $\widehat{e}_i(\bD_{N_i}, z_i)$ and $\widehat{e}'_{ij}(D_j, \bD_{N_j}^{-i}, z_i, z_j)$. To improve robustness against model misspecification, we propose doubly robust (DR) estimators that combine the IPW approach with outcome regression models. The key advantage of DR estimators is that they remain consistent if either the propensity score model or the outcome regression model is correctly specified, providing protection against misspecification of one of these components.

To construct the DR estimators, we first define outcome regression models that align with our identification assumptions. For ADTT, we define the conditional expectation functions:
\begin{align*}
\mu_{1i}(\bD_{N_i}, z_i) &= E[Y_{2i} - Y_{1i} | D_i=1, \bD_{N_i}, z_i], \\
\mu_{0i}(\bD_{N_i}, z_i) &= E[Y_{2i} - Y_{1i} | D_i=0, \bD_{N_i}, z_i].
\end{align*}
These functions condition on the full neighborhood treatment vector $\bD_{N_i}$, ensuring consistency with Assumption~\ref{as:trend}. For AITT, we define similar outcome regression models for neighbor $j$'s outcomes, conditioning on $D_i$, $D_j$, and $\bD_{N_j}^{-i}$:
\begin{align*}
\mu_{1ij}(D_j, \bD_{N_j}^{-i}, z_i, z_j) &= E[Y_{2j} - Y_{1j} | D_i=1, D_j, \bD_{N_j}^{-i}, z_i, z_j], \\
\mu_{0ij}(D_j, \bD_{N_j}^{-i}, z_i, z_j) &= E[Y_{2j} - Y_{1j} | D_i=0, D_j, \bD_{N_j}^{-i}, z_i, z_j].
\end{align*}
Based on these outcome regression models, we propose the following DR estimators. Here, we denote the observed outcome difference as $\Delta Y_i = Y_{2i} - Y_{1i}$ (and similarly for $\Delta Y_j$). The predicted outcome differences are defined as:
\begin{align*}
\Delta m_{1i} &= \widehat{\mu}_{1i}(\bD_{N_i}, z_i), \quad 
\Delta m_{0i} = \widehat{\mu}_{0i}(\bD_{N_i}, z_i), \\
\Delta m'_{1,ij} &= \widehat{\mu}_{1ij}(D_j, \bD_{N_j}^{-i}, z_i, z_j), \quad 
\Delta m'_{0,ij} = \widehat{\mu}_{0ij}(D_j, \bD_{N_j}^{-i}, z_i, z_j),
\end{align*}
where $\widehat{\mu}_{1i}$, $\widehat{\mu}_{0i}$, $\widehat{\mu}_{1ij}$, and $\widehat{\mu}_{0ij}$ are the estimators of the corresponding outcome regression functions defined above.

We estimate outcome regression models using linear regression. For ADTT, we regress $\Delta Y_i$ on $D_i$, $z_i$, and the distance-sorted neighbor treatments $\bD_{N_i}$. Predictions $\Delta m_{1i}$ and $\Delta m_{0i}$ are obtained by setting $D_i=1$ and $D_i=0$, respectively.
For AITT, we regress $\Delta Y_j$ on $D_i$, $D_j$, $z_i$, $z_j$, and the sorted neighbor treatments $\bD_{N_j}^{-i}$. Predictions are generated by fixing $D_i$. While we use OLS, flexible methods like machine learning can also be employed.
\begin{equation}\label{eq:DR_ADTT}
\begin{split}
\widehat{\tau}_{\mathrm{dir}}^{dr} & = \frac{1}{n} \sum_{i = 1}^n \Bigg\{\frac{D_i}{\widehat{\pi}_i}\left(\Delta Y_i-\Delta m_{1i}\right) 
-\frac{(1-D_i)\widehat{e}_i}{\widehat{\pi}_i(1-\widehat{e}_i)}\left(\Delta Y_i-\Delta m_{0i}\right) + \frac{\widehat{e}_i}{\widehat{\pi}_i}\left(\Delta m_{1i}-\Delta m_{0i} \right)\Bigg\},
\end{split}
\end{equation}
and for AITT,
\begin{equation}\label{eq:DR_AITT}
\begin{split}
\widehat{\tau}_{\mathrm{ind}}^{dr} &= \frac{1}{n}\sum_{i=1}^n \Bigg\{\frac{1}{|N_i|}\sum_{j\in N_i}
\Bigg(\frac{D_i}{\widehat{\pi}_i}\left( \Delta Y_j-\Delta m^{'}_{1,ij}\right)\\
& -\frac{(1-D_i)\widehat{e'}_{ij}}{\widehat{\pi}_i(1-\widehat{e'}_{ij})}\left(\Delta Y_j-\Delta m^{'}_{0,ij}\right) + \frac{\widehat{e'}_{ij}}{\widehat{\pi}_i}(\Delta m^{'}_{1,ij}-\Delta m^{'}_{0,ij})\Bigg) \Bigg\}.
\end{split}
\end{equation}

Structurally, these estimators consist of the standard IPW component augmented by bias-correction terms derived from the outcome regression models. If the propensity score is misspecified, these regression terms correct the residual bias, and conversely, the IPW weights ensure consistency if the outcome model is incorrect. Here, $\widehat{e}_i$ and $\widehat{e}'_{ij}$ are shorthand notations for $\widehat{e}_i(\bD_{N_i}, z_i)$ and $\widehat{e}'_{ij}(D_j, \bD_{N_j}^{-i}, z_i, z_j)$, respectively. Thus, the DR estimators possess the double robustness property that they are consistent for ADTT and AITT if either the propensity score models ($\widehat{e}_i$, $\widehat{e}'_{ij}$) or the outcome regression models ($\widehat{\mu}_{1i}$, $\widehat{\mu}_{0i}$, $\widehat{\mu}_{1ij}$, $\widehat{\mu}_{0ij}$) are correctly specified. This property provides substantial protection against model misspecification, as researchers need only correctly specify one of the two model components rather than both. When both models are correctly specified, the DR estimators typically achieve improved efficiency compared to the IPW estimators alone.

\subsection{Variance estimation}
To conduct statistical inference on ADTT and AITT, we need an estimator for the (asymptotic) variance of $\widehat{\tau}_{\mathrm{dir}}$ and $\widehat{\tau}_{\mathrm{ind}}$. 
Under network dependence, standard variance estimators are inconsistent due to the correlation among observations. This issue is particularly pronounced in applications, such as those analyzed by \cite{conley1999gmm}, where geographically proximate agents or firms within the same industry are subject to common shocks, leading to correlated disturbances that standard estimators do not account for.
Therefore, we employ a heteroskedasticity and autocorrelation consistent (HAC) estimator that accounts for the network dependence structure, following \citet{Kojevnikov2021}.
Specifically, we consider the following kernel-based HAC estimator:
\begin{equation}\label{eq:HAC}
\widehat{V}_n = \sum_{s \geq 0} \omega\left(\frac{s}{b_n}\right) \widehat{\Omega}_n(s),
\end{equation}
where $\omega$ is a kernel function satisfying $\omega(0) = 1$, having a finite support, and being symmetric. 
Examples include Bartlett and Parzen kernels. 
The term $b_n$ is a bandwidth parameter that increases with $n$. In practice, the bandwidth $b_n$ is typically selected based on the network structure and the interference range $K$. Following the framework of \citet{Kojevnikov2021}, we set the bandwidth as a multiple of the interference range $K$ (e.g., $b_n = c \cdot K$ for some constant $c > 0$), which accounts for the local dependence structure on the network. 
$\widehat{\Omega}_n(s)$ is the sample autocovariance matrix at distance $s$, given by:
$$
\widehat{\Omega}_n(s) = n^{-1} \sum_{i \in I_n} \sum_{j \in N^{\partial}(i; s)} (\phi_i - \bar{\phi}_n)(\phi_j - \bar{\phi}_n)
$$
where $N^{\partial}(i; s)$ is the set of units at exactly distance $s$ from unit $i$ (as defined in Section 2), and $\phi_i$ is the influence function of the estimand as follows:
\begin{equation}\label{eq:Z}
\begin{split}
{\rm (ADTT)} \ \ \ &
\phi_i^{\mathrm{dir}} = \frac{D_i - e_i(\bD_{N_i}, z_i)}{\pi_i(z_i) \{1-e_i(\bD_{N_i}, z_i)\}} ( Y_{2i} - Y_{1i} ),\\
{\rm (AITT)} \ \ \ &
\phi_i^{\mathrm{ind}} = \frac{1}{|N_i|} \sum_{j \in N_i} \left\{ \frac{D_i - e'_{ij}(D_j, \bD_{N_j}^{-i}, z_i, z_j)}{\pi_i(z_i)\{1-e'_{ij}(D_j, \bD_{N_j}^{-i}, z_i, z_j)\}} (Y_{2j} - Y_{1j}) \right\}.
\end{split}
\end{equation}
Here $\bar{\phi}_n$ is its sample mean.

When using the DR estimators introduced in Section 3.2, the influence function $\phi_i$ should be replaced with the corresponding summand from the DR estimator. Specifically, for the DR estimators, the influence functions are:
\begin{equation}\label{eq:Z_DR}
\begin{split}
{\rm (ADTT)} \ \ \ &
\phi_i^{DR,\mathrm{dir}} = \frac{D_i}{\widehat{\pi}_i}\left(\Delta Y_i-\Delta m_{1i}\right) -\frac{(1-D_i)\widehat{e}_i}{\widehat{\pi}_i(1-\widehat{e}_i)}\left(\Delta Y_i-\Delta m_{0i}\right) + \frac{\widehat{e}_i}{\widehat{\pi}_i}\left(\Delta m_{1i}-\Delta m_{0i} \right),\\
{\rm (AITT)} \ \ \ &
\phi_i^{DR,\mathrm{ind}} = \frac{1}{|N_i|} \sum_{j \in N_i} \Bigg\{ \frac{D_i}{\widehat{\pi}_i}\left( \Delta Y_j-\Delta m^{'}_{1,ij}\right) \\
& -\frac{(1-D_i)\widehat{e'}_{ij}}{\widehat{\pi}_i(1-\widehat{e'}_{ij})}\left(\Delta Y_j-\Delta m^{'}_{0,ij}\right) + \frac{\widehat{e'}_{ij}}{\widehat{\pi}_i}(\Delta m^{'}_{1,ij}-\Delta m^{'}_{0,ij}) \Bigg\}.
\end{split}
\end{equation}

This estimator robustly captures the local dependence structure by accounting for the covariance between units that are close on the network and down-weighting pairs of units as the distance between them increases. As shown in the subsequent section, the HAC estimator $\widehat{V}_n$ is consistent as $n \to \infty$. 
Based on the variance estimator, one can make statistical inference on ADTT or AITT.
For example, the Wald-type $100(1-\alpha)\%$ confidence interval of ADTT can be constructed as $\widehat{\tau}_{\mathrm{dir}}\pm z_{\alpha/2}\sqrt{\widehat{V}_n/n}$.

\section{Asymptotic Properties}

This section establishes the large-sample properties (consistency and asymptotic normality) of the proposed estimators. Since data on a network structure exhibit complex dependencies, standard asymptotic theory based on the i.i.d. assumption is not applicable. We therefore employ modern theory for weakly dependent random fields, specifically the framework of $\psi$-dependence introduced by \citet{Kojevnikov2021}, to provide rigorous proofs. This approach is also related to the analysis of approximate neighborhood interference by \citet{Leung2022}.

\subsection{Local dependence on networks}

First, we define the concept of weak dependence on which our analysis relies. $\psi$-dependence is a measure of the strength of dependence based on the covariance of non-linearly transformed variables.

\begin{df}[$\psi$-dependence]
A collection of random variables $\{\phi_i\}_{i=1 \ldots n}$ is said to be $\psi$-dependent if there exist a set of non-negative functions $\{\psi_{h,h'}(\cdot,\cdot)\}$ and a uniformly bounded sequence $\{\tilde{\theta}_{n, s}\}$ with $\tilde{\theta}_{n, 0} = 1$ such that for any $n, h, h', s > 0$, any bounded Lipschitz functions $f \in L_h, f' \in L_{h'}$, and any pair of sets $(H, H') \in \mathcal{P}_n(h, h', s)$ separated by a network distance of at least $s$, the following holds:
$$
|\text{Cov}(f(\phi_H), f'(Z_{H'}))| \leq \psi_{h,h'}(f,f') \tilde{\theta}_{n, s}
$$
where $\mathcal{P}_n(h, h', s) = \{(H, H') \mid l_A(H, H') \geq s, |H| = h, |H'| = h' \}$, $\mathcal{L}_{a}$ denotes the collection of bounded Lipschitz real functions on $\mathbb{R}^{a}$, i.e., $\mathcal{L}_{a} = \{ f : \mathbb{R}^{a} \to \mathbb{R} : \|f\|_{\infty} < \infty, \ \mathrm{Lip}(f) < \infty \}$ with $\mathrm{Lip}(f)$ denoting the Lipschitz constant of $f$,$^6$ and $\|\cdot\|_{\infty}$ the sup-norm of $f$, i.e., $\|f\|_{\infty} = \sup_x |f(x)|.$ 
\end{df}

\noindent
This definition requires the dependence coefficient $\tilde{\theta}_{n, s}$ to decay as the distance $s$ increases.

\subsection{Assumptions}

To derive the large-sample properties of our estimators, we introduce a set of regularity conditions regarding the scope of interference, stochastic properties, and the overall dependence structure of the network.

\begin{as}[Local interference]
\label{as:local_interference}
Interference is limited to a distance $K$. That is, the fundamental variables for unit $i$, $(Y_{1i}, D_i, \epsilon_i)$, do not depend on the treatment of units beyond distance $K$. For all $i \in I_n$ and for all $\bD_{-i} \in \{0, 1\}^{n-1}$:
$$
P(Y_{1i}, D_i, \epsilon_i | \bD_{-i} = \bd_{-i}) = P(Y_{1i}, D_i, \epsilon_i | \bD_{N(i; K)} = \bd_{N(i; K)}).
$$
\end{as}

\begin{as}[Overlap]
\label{as:overlap}
All propensity scores are uniformly bounded away from 0 and 1. That is, there exist $\underline{\pi}, \overline{\pi} \in (0, 1)$ such that for all $i, j \in I_n$, $\pi_i(z_i) = P(D_i = 1 | z_i) \in [ \underline{\pi}, \overline{\pi} ]$, $e_i(\bD_{N_i}, z_i) = P(D_i = 1 | \bD_{N_i}, z_i) \in [ \underline{\pi}, \overline{\pi} ]$ and $e'_{ij}(D_j, \bD_{N_j}^{-i}, z_i, z_j) = P(D_i = 1 | D_j, \bD_{N_j}^{-i}, z_i, z_j) \in [ \underline{\pi}, \overline{\pi} ]$.
\end{as}

\begin{as}[Bounded outcomes]
\label{as:bounded_outcomes}
Outcomes are uniformly bounded. That is, there exists a constant $\overline{Y} < \infty$ such that for all $i \in I_n$, $|Y_{1i}|, |Y_{2i}| < \overline{Y}$ almost surely.
\end{as}

Next, we introduce an assumption about the dependence structure on the network. Let $x_i = (Y_{1i}, D_i, \epsilon_i)$. We assume that the collection of these fundamental random variables $\{x_i\}$ is spatially weakly dependent. Specifically, we assume a spatial $\alpha$-mixing condition of the type considered in \citet{Jenish2009}. Limit theorems for such dependence structures have been developed in a series of papers by the same authors (e.g., \citealp{Jenish2009, Jenish2012}).

\begin{as}[$\alpha$-mixing]
\label{as:alpha_mixing}\
Let $\alpha_n(H, H') = \alpha(\sigma_n(\{x_i\}_{i \in H}), \sigma_n(\{x_i\}_{i \in H'}))$, where $\sigma(\cdot)$ denotes the $\sigma$-algebra generated by the collection of random variables. There exist a bounded function $\psi(\cdot, \cdot): \mathbb{N} \times \mathbb{N} \rightarrow \mathbb{R}$ and a sequence $\{\widehat{\alpha}(s)\}_{s \in \mathbb{R}}$ such that for $\bar{\alpha}(h, h', s) = \sup_n \sup_{H, H'} \{ \alpha_n(H, H') \mid |H| \leq h, |H'| \leq h', l_A(H, H') \geq s \}$, we have:
$$
\bar{\alpha}(h, h', s) \leq \psi(h, h') \widehat{\alpha}(s)
$$
where $\widehat{\alpha}(s) \to 0$ as $s \to \infty$. This assumption governs the strength of the underlying spatial dependence in the observed data.
\end{as}

\noindent
The following technical assumptions specify the trade-off between this decay rate and the network's density, ensuring the Law of Large Numbers (LLN) and the Central Limit Theorem (CLT) hold.

\begin{as}[Weak dependence for LLN]
\label{as:LLN}
$\sum_{s=0}^{n} M_n^{\partial}(s) \tilde{\theta}_{n, s} = o(n)$, where $M_n^{\partial}(s) = \frac{1}{n} \sum_{i=1}^n |N^{\partial}(i; s)|$ and $N^{\partial}(i; s)$ is as defined in Section 2.
\end{as}

\begin{as}[Weak dependence for CLT]
\label{as:CLT}
A stronger technical condition on the network structure and dependence decay rate is required for the CLT. There exists a positive sequence $r_n \to \infty$ such that for some $p > 4$ and $k=1,2$:
$$
\frac{1}{\sigma_n^{2+k}} \sum_{i \in I_n} \sum_{s=0}^{\infty} s^{d-1} \max_{j \in I_n, s \leq l_A(i, j) < s+1} |N(i; r_n)^+ \setminus N(j; s-1)^+|^k \tilde{\theta}_{n, s}^{1-\frac{2+k}{p}} \to 0
$$
and
$$
\frac{|I_n|^2 \tilde{\theta}_{n, r_n}^{1-\frac{1}{p}}}{\sigma_n} \to 0 \quad \text{as } n \to \infty
$$
where $\sigma_n^2$ is the variance of the sum of the random variables.
\end{as}

\noindent
Finally, we include the assumptions required for the consistency of the HAC variance estimator, based on Proposition 4.1 in \citet{Kojevnikov2021}.

\begin{as}[Conditions for HAC consistency]
\label{as:HAC}
There exists some $p > 4$ such that the following conditions on the kernel $\omega(\cdot)$, the bandwidth $b_n$, and the weak dependence coefficients $\tilde{\theta}_{n, s}$ are satisfied:
\begin{itemize}
\item 
$\omega(0)=1, \omega(s/b_n)=0 \text { for any } s>b_n$ and $\left|\omega(s/b_n)\right|<\infty$ for all $n$.
\item 
$\lim_{n \rightarrow \infty} \sum_{s \geq 1}\left|\omega\left(s/b_n\right)-1\right| s^{d-1} \,\theta_{n, s}^{1-\frac{2}{p}}=0$.
\item $\lim_{n \rightarrow \infty} n^{-1}\sum_{s=0}^{\infty} s^{d-1} \max _{j \in I_n, s \leq l_A(i, j)<s+1}\left|N\left(i ; b_n \right)^+ \setminus N(j ; s-1)^+\right|^2 \,\theta_{n, s}^{1- \frac{4}{p}}=0$.
\item 
$\delta_n\left(b_n\right)=o(n)$ almost surely, where $\delta_n(b_n) = \sum_{s=0}^{b_n} M_n^{\partial}(s)$ and $M_n^{\partial}(s) = \frac{1}{n} \sum_{i=1}^n |N^{\partial}(i; s)|$ is the average number of units at exactly distance $s$ from each unit. This condition ensures that the network density within the bandwidth $b_n$ does not grow too fast relative to the sample size, which is necessary for the consistency of the HAC estimator.
\item
The sequence $\left\{\tilde{\theta}_{n, s} / s^{p /\left(p-4\right)}\right\}$ is non-increasing in $s \geq 1$.
\end{itemize}
\end{as}

\subsection{Main theorems}

First, we present lemmas showing that the components of our proposed estimators satisfy the definition of $\psi$-dependence. This is derived from the $\alpha$-mixing property of the underlying variables (Assumption \ref{as:alpha_mixing}) and local interference (Assumption \ref{as:local_interference}). The proofs are provided in the Appendix. Throughout this subsection, $\mathbf{1}\{\cdot\}$ denotes the indicator function, $\mathbf{1}\{A\} = 1$ if condition $A$ holds and $\mathbf{1}\{A\} = 0$ otherwise.

\begin{lem}[$\psi$-dependence of ADTT and AITT summand]
\label{lem:psi_ADTT}
Under Assumptions \ref{as:local_interference}-\ref{as:alpha_mixing}, the collection of random variables $\{ \phi_i^{\mathrm{dir}} \}_{i \in I_n}$ and $\{ \phi_i^{\mathrm{ind}} \}_{i \in I_n}$ defined in (\ref{eq:Z}) are $\psi$-dependent with $\tilde{\theta}_{n, s} = \mathbf{1}\{s < 2K\} + \mathbf{1}\{s \geq 2K\} \widehat{\alpha}(s-2K)$ and $\tilde{\theta}_{n, s} = \mathbf{1}\{s < 4K\} + \mathbf{1}\{s \geq 4K\} \widehat{\alpha}(s-4K)$, respectively. 
\end{lem}

This lemma ensures that our estimators can be analyzed within the $\psi$-dependence framework. By applying the limit theorems from \citet{Kojevnikov2021}, we obtain consistency and asymptotic normality.

\begin{thm}[Consistency]
\label{thm:consistency_ADTT}
Under Assumptions \ref{as:trend} and \ref{as:local_interference}-\ref{as:LLN}, $\widehat{\tau}_{\mathrm{dir}}$ and $\widehat{\tau}_{\mathrm{ind}}$ are consistent under $n\to\infty$. 
\end{thm}

\begin{thm}[Asymptotic normality]
\label{thm:normality_ADTT}
Under Assumptions \ref{as:trend}, \ref{as:local_interference}-\ref{as:alpha_mixing}, and \ref{as:CLT}, it holds that $\sqrt{n}(\widehat{\tau}_{\mathrm{dir}}-\tau_{\mathrm{dir}})/\sigma_{n,D} \to N(0, 1)$ and $\sqrt{n}(\widehat{\tau}_{\mathrm{ind}}-\tau_{\mathrm{ind}})/\sigma_{n,I} \to N(0, 1)$ as $n\to\infty$, where $\sigma_{n,D}^2={\rm Var}(n^{-1}\sum_{i \in I_n}\phi_i^{\mathrm{dir}})$ and $\sigma_{n,I}^2={\rm Var}(n^{-1}\sum_{i \in I_n}\phi_i^{\mathrm{ind}})$.

\end{thm}

We next provide consistency of the HAC estimator introduced in Section~3.3. 

\begin{thm}[Consistency of variance estimator]
\label{thm:HAC}\
Under Assumptions \ref{as:trend}, \ref{as:local_interference}-\ref{as:alpha_mixing}, \ref{as:CLT} and \ref{as:HAC}, the HAC estimator $\widehat{V}_n$ given in (\ref{eq:HAC}) is consistent for the asymptotic variance. Specifically, for the ADTT estimator, $\widehat{V}_n \to \sigma_{n,D}^2$ as $n\to\infty$, and for the AITT estimator, $\widehat{V}_n \to \sigma_{n,I}^2$ as $n\to\infty$, where $\sigma_{n,D}^2$ and $\sigma_{n,I}^2$ are defined in Theorem \ref{thm:normality_ADTT}.
\end{thm}

We now provide theoretical guarantees for the DR estimators. The first theorem establishes the double robustness property, which is the key advantage of DR estimators over IPW estimators.

\begin{thm}[Double robustness of DR estimators]
\label{thm:DR_robustness}
Under Assumptions \ref{as:NoAnticipation} and \ref{as:trend}, the DR estimators $\widehat{\tau}_{\mathrm{dir}}^{DR}$ and $\widehat{\tau}_{\mathrm{ind}}^{DR}$ are consistent for $\tau_{\mathrm{dir}}$ and $\tau_{\mathrm{ind}}$, respectively, if either the propensity score models ($e_i$, $e'_{ij}$) are correctly specified, or the outcome regression models ($\mu_{1i}$, $\mu_{0i}$, $\mu_{1ij}$, $\mu_{0ij}$) are correctly specified.
\end{thm}

\begin{thm}[Asymptotic normality of DR estimators]
\label{thm:DR_normality}
Under Assumptions \ref{as:trend}, \ref{as:local_interference}-\ref{as:alpha_mixing}, and \ref{as:CLT}, if both the propensity score models and the outcome regression models are correctly specified (or estimated using machine learning methods with sufficient convergence rates), then $\sqrt{n}(\widehat{\tau}_{\mathrm{dir}}^{DR}-\tau_{\mathrm{dir}})/\sigma_{n,D}^{DR} \to N(0, 1)$ and $\sqrt{n}(\widehat{\tau}_{\mathrm{ind}}^{DR}-\tau_{\mathrm{ind}})/\sigma_{n,I}^{DR} \to N(0, 1)$ as $n\to\infty$, where $\sigma_{n,D}^{DR,2}={\rm Var}(n^{-1}\sum_{i \in I_n}\phi_i^{DR,\mathrm{dir}})$ and $\sigma_{n,I}^{DR,2}={\rm Var}(n^{-1}\sum_{i \in I_n}\phi_i^{DR,\mathrm{ind}})$ with $\phi_i^{DR,\mathrm{dir}}$ and $\phi_i^{DR,\mathrm{ind}}$ defined in equation (\ref{eq:Z_DR}).
\end{thm}

All the proofs of the above theorems are provided in the Appendix. 
The above theorems provide theoretical guarantees of the statistical inference on ADTT and AITT by using the HAC estimator $\widehat{V}_n$.

\section{Simulation Study}

This subsection evaluates the finite-sample performance of our proposed estimators (ADTT and AITT) through a series of Monte Carlo simulations. The primary goal is to demonstrate the robustness of our identification-based approach in settings where the true exposure mapping is unknown and potentially complex, comparing it against existing methods that rely on pre-specified structures.

\subsection{Data generating process}

\begin{figure}[!htbp]
\centering
\includegraphics[width=0.9\linewidth]{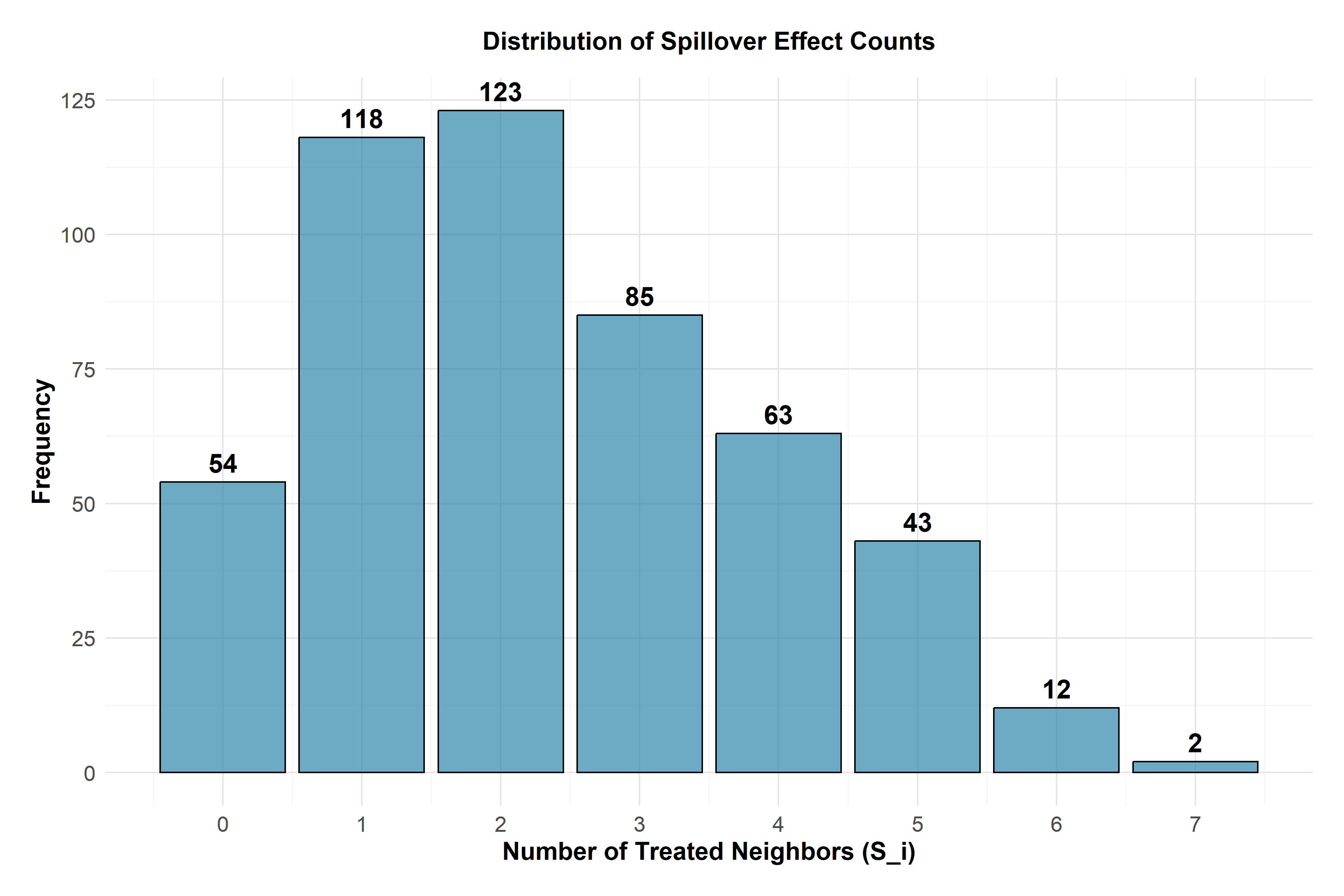}
\caption{Distribution of the number of treated neighbors ($S_i$) across units in a representative simulation run. This histogram shows the prevalence of different exposure levels, which is crucial for understanding the distribution of spillover effects in our data generating process.}
\label{fig:spillover_structure}
\end{figure}

We simulate a population of $n$ units randomly located in a $20 \times 20$ two-dimensional space. The network structure is defined by the Chebyshev distance with the interference range set to $K=1$.
We generate two types of confounder, an observed individual attribute $z_i \sim N(0,1)$, and an unobserved dependent factor, $(z_{u1},\ldots,z_{un}) \sim N(0, \Sigma)$, where the covariance matrix $\Sigma_{ij} = 0.5^{l_A(i, j)}$ captures the network dependence that decays with the distance of the network $l_A(i, j)$. 
Then, the treatment is assigned based on both factors as 
$$
P(D_i=1 \mid z_i, z_{u,i}) = \text{logit}^{-1}(0.3 z_i + 0.8 z_{u,i}),
$$
which generates a network-dependent clustered treatment pattern.
The outcomes are generated according to the following two-period model:
\begin{align*}
Y_{1i} &= 1.2 z_i + 0.5 z_{u,i} + \epsilon_{1i} \\
Y_{2i} &= 1 + Y_{1i} + \tau D_i + f(S_i) + 0.1 z_{u,i} + 0.2 z_i + \epsilon_{2i}
\end{align*}
where $\epsilon_{1i}, \epsilon_{2i} \sim N(0, 1)$. The true direct treatment effect is set to $\tau = 0.8$.
The core feature of this simulation is the spillover function $f(S_i)$, where $S_i$ is the number of treated neighbors of unit $i$. We assume a complex, step-wise increasing structure:
$$
f(S_i)=0.8 \cdot I(S_i=1) + 1.6\cdot I(S_i=2) + 2.4 \cdot I(S_i\geq 3).
$$
This represents a non-linear interference structure where the spillover effect saturates. We assume that this functional form $f(\cdot)$ is unknown to the researcher.
While the true ADTT is equal to $\tau=0.8$, the true AITT is calculated as the conditional average over the generated data. Specifically, for each treated unit $i$, we compute the average spillover effect on its neighbors $j$ by comparing the outcomes when unit $i$ is treated versus not treated, conditional on the observed network structure and treatment assignments. The true AITT calculated on the basis of the DGP, averaged across simulation runs, is approximately 0.4961. 
Figure \ref{fig:spillover_structure} shows the distribution of $S_i$ in a representative simulation run, indicating the prevalence of different levels of exposure.

\subsection{Comparative methods}

We first evaluate the performance of the estimation of the direct effect of the treatment, obtained by the proposed estimators (which rely on nonparametric identification) together with several existing methods. 
For the proposed estimators, propensity scores are estimated conditioning on the observed covariates $z_i$ and the neighborhood treatment vector $\bD_{N_i}$ (for ADTT) or $\bD_{N_j}^{-i}$ (for AITT), as specified in Theorems \ref{thm:identification_ADTT} and \ref{thm:identification_AITT}. Following Section 3.1, we estimate propensity scores using logistic regression. In our simulation experiments, we set $L = 10$. For AITT estimation, when the number of neighbors for unit $j$ exceeds $L$, we randomly sample $L$ neighbors from those within distance $K$ to maintain computational efficiency while avoiding arbitrary selection.

All other estimators use only the observed covariates $z_i$.
In this simulation, we report results for both the IPW and DR versions of our proposed estimators. For the DR estimators, we estimate outcome regression models using linear regression (OLS) as described in Section 3.2.

\begin{itemize}
\item 
\textbf{Proposed IPW}: Our proposed nonparametric IPW estimators for ADTT and AITT, implemented as described in Section 3.1 with $L=10$.
\item
\textbf{Proposed DR}: Our proposed nonparametric DR estimators for ADTT and AITT, implemented as described in Section 3.2.

\item
\textbf{Xu (2025) methods}: The estimators proposed by \citet{Xu2025}, which require specifying an exposure mapping $G_i$. We implement both the IPW and DR versions of these estimators. The following three exposure mappings are used:
\begin{itemize}
\item 
Oracle (Infeasible): The researcher knows the true 4-level structure. $G_i^{\text{Oracle}} = \min(S_i, 3)$.
\item 
MO (Misspecified Ordering): The structure is misspecified; 30\% of units are randomly assigned to a different exposure group based on the Oracle exposure mapping $G_i^{\text{Oracle}}$.
\item 
FM (Fully Misspecified): The researcher assumes an overly simplified binary structure. $G_i^{FM} = \mathbb{I}(S_i > 1)$.
\end{itemize}
Note that in Table \ref{tab:main_results_adtt}, we report only the DR versions (denoted as Xu (Oracle), Xu (MO), and Xu (FM)) for comparison with our proposed DR estimator. Figure \ref{fig:estimator_distribution} includes both IPW and DR versions of Xu's methods.

\item 
\textbf{Standard DID methods}: Standard DID estimators that ignore interference (canonical IPW, canonical TWFE, DR-DID) and an extended TWFE model (Modified TWFE in \cite{Xu2025}) that accounts for simple binary exposure ($I(S_i \geq 1)$).
\end{itemize}

\subsection{Results}
\begin{figure}[!htbp]
\centering
\includegraphics[height=6cm]{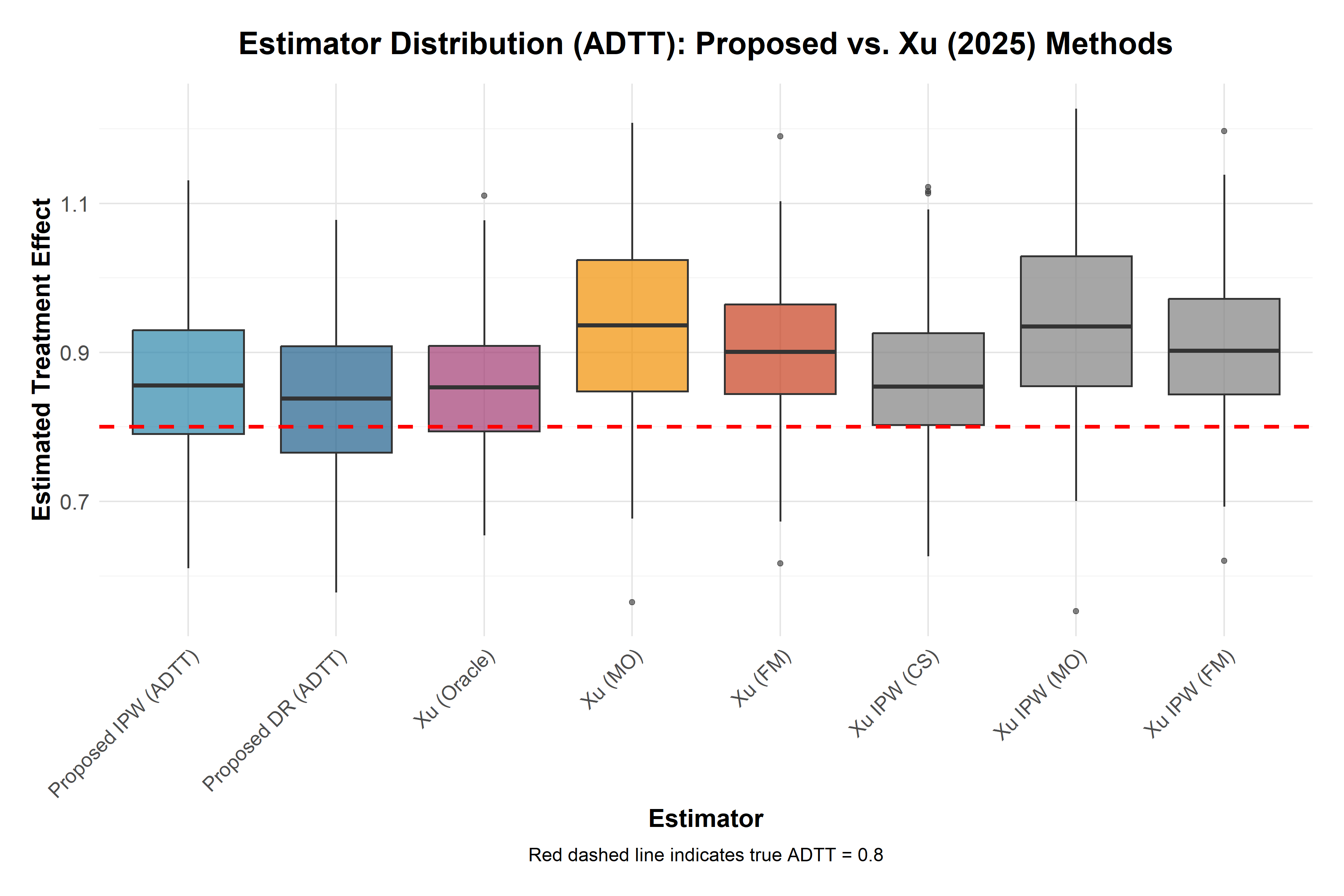}
\hfill
\includegraphics[height=6cm]{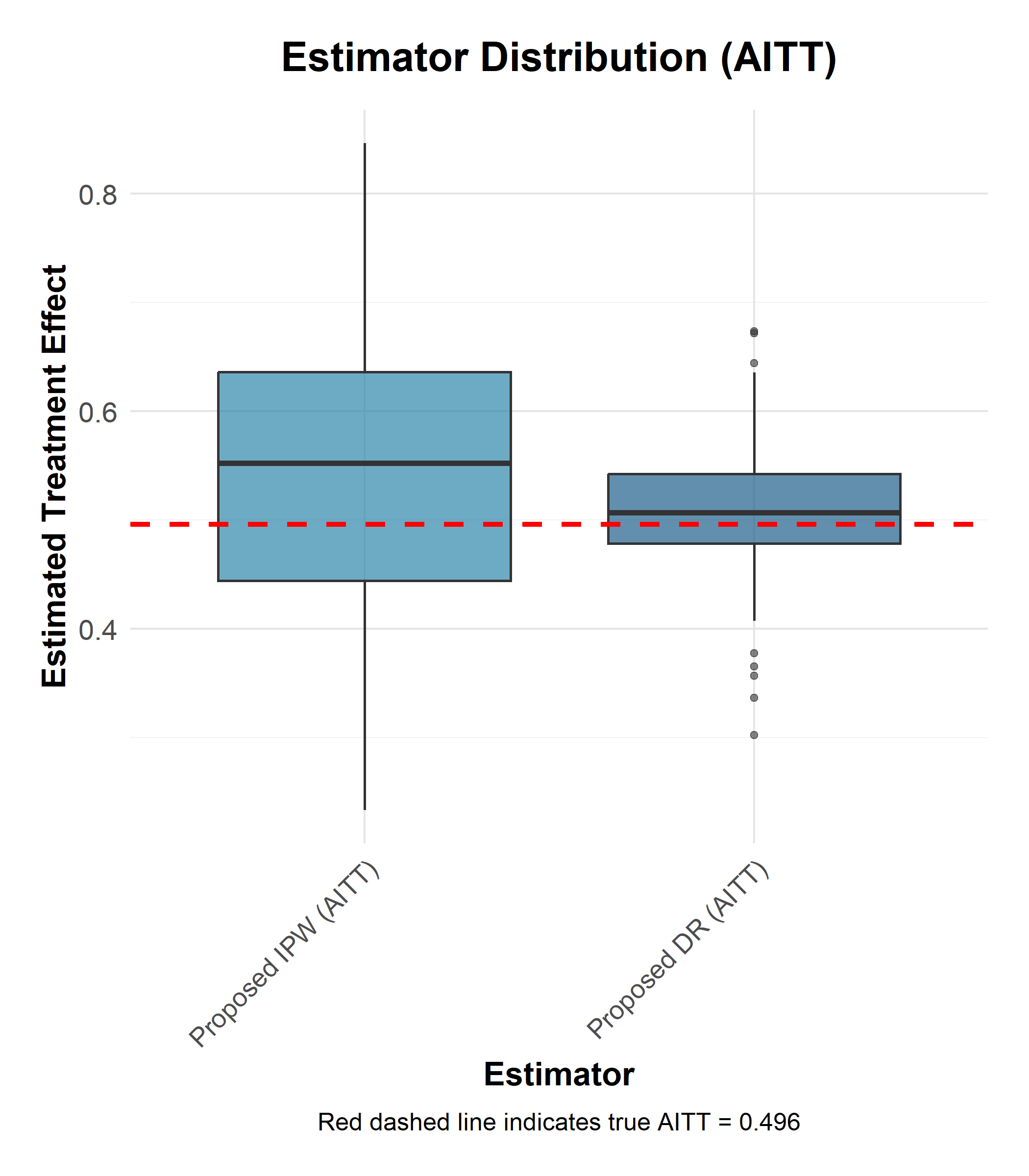}

\caption{Distribution of estimated treatment effects across Monte Carlo replications. Left: ADTT estimates. Right: AITT estimates. True values indicated by dashed lines (ADTT = 0.8, AITT = 0.4961). Methods are labeled as Proposed IPW/DR (our methods), Xu (Oracle), Xu (MO), and Xu (FM) corresponding to the exposure mapping specifications described in Section 5.2.}
\label{fig:estimator_distribution}
\end{figure}

Table \ref{tab:main_results_adtt} and Figure \ref{fig:estimator_distribution} summarize the results for ADTT ($\tau_{\mathrm{dir}} = 0.8$). The results show that the proposed IPW and DR estimators tend to achieve substantially lower bias and RMSE than methods that rely on misspecified exposure mappings or ignore interference. In particular, the proposed ADTT estimators have bias (0.062 and 0.033, respectively) and RMSE comparable to the infeasible Oracle estimator. This is likely because conditioning on the full neighborhood treatment vector $\bD_{N_i}$ avoids exposure mapping misspecification while the conditional parallel trends assumption remains valid. Benchmark estimators ignoring interference (Canonical IPW, TWFE, DR-DID) exhibit substantial upward bias (Bias $\approx 0.21$) due to positive spillovers from clustered treatments; methods from \citet{Xu2025} with misspecified exposure mappings (MO, FM) also show significant bias.

\begin{table}[h]
\centering
\label{tab:main_results_adtt}
\begin{tabular}{lrrr}
\hline
Estimator & Bias & RMSE & CP \\
\hline
\multicolumn{4}{l}{\textit{Proposed Methods}} \\
\textbf{Proposed IPW (ADTT)} & \textbf{0.0619} & \textbf{0.1171} & \textbf{1.0000} \\
\textbf{Proposed DR (ADTT)} & \textbf{0.0331} & \textbf{0.1040} & \textbf{0.9500} \\
\hline
\multicolumn{4}{l}{\textit{Xu (2025) Methods}} \\
Xu (Oracle) & 0.0567 & 0.1064 & 0.9900 \\
Xu (MO) & 0.1358 & 0.1830 & 0.9400 \\
Xu (FM) & 0.1032 & 0.1411 & 0.9900 \\
\hline
\multicolumn{4}{l}{\textit{Standard DID Methods}} \\
Canonical IPW & 0.2115 & 0.2621 & 0.5200 \\
Canonical TWFE & 0.2077 & 0.2591 & 0.5100 \\
DR-DID & 0.2100 & 0.2613 & 0.5200 \\
Modified TWFE & 0.1484 & 0.1936 & 0.7000 \\
\hline
\end{tabular}
\caption{Simulation Results for ADTT under $n=500$ (100 replications). True ADTT = 0.8.
}
\end{table}

\begin{table}[h]
\centering
\label{tab:main_results_aitt}
\begin{tabular}{lrrr}
\hline
Estimator & Bias & RMSE & CP \\
\hline
\multicolumn{4}{l}{\textit{Proposed Methods}} \\
\textbf{Proposed IPW (AITT)} & \textbf{0.0473} & \textbf{0.1237} & \textbf{1.0000} \\
\textbf{Proposed DR (AITT)} & \textbf{0.0139} & \textbf{0.0575} & \textbf{0.9200} \\
\hline
\end{tabular}
\caption{Simulation Results for AITT: Performance comparison under $n=500$ based on 100 Monte Carlo replications. Bias, RMSE, and Coverage Probability (CP) at 95\% confidence level are reported. The true AITT is 0.4961. See Section 5.2 for details on each estimator.
}
\end{table}

The proposed AITT estimators demonstrate strong performance. In particular, the DR AITT estimator has the smallest bias (0.0139) and RMSE (0.0575) among feasible estimators; the IPW AITT estimator achieves bias 0.0473 and RMSE 0.1237, comparable to or better than the corresponding ADTT estimators. This is likely because conditioning on the neighborhood treatment configuration for both unit $i$ and its neighbors $j$ preserves identification without exposure mapping, while the curse of dimensionality remains manageable at moderate neighborhood sizes. The DR AITT estimator's superior performance illustrates the benefit of double robustness for indirect effect estimation.

\begin{figure}[!htbp]
\includegraphics[width=1.0\linewidth]{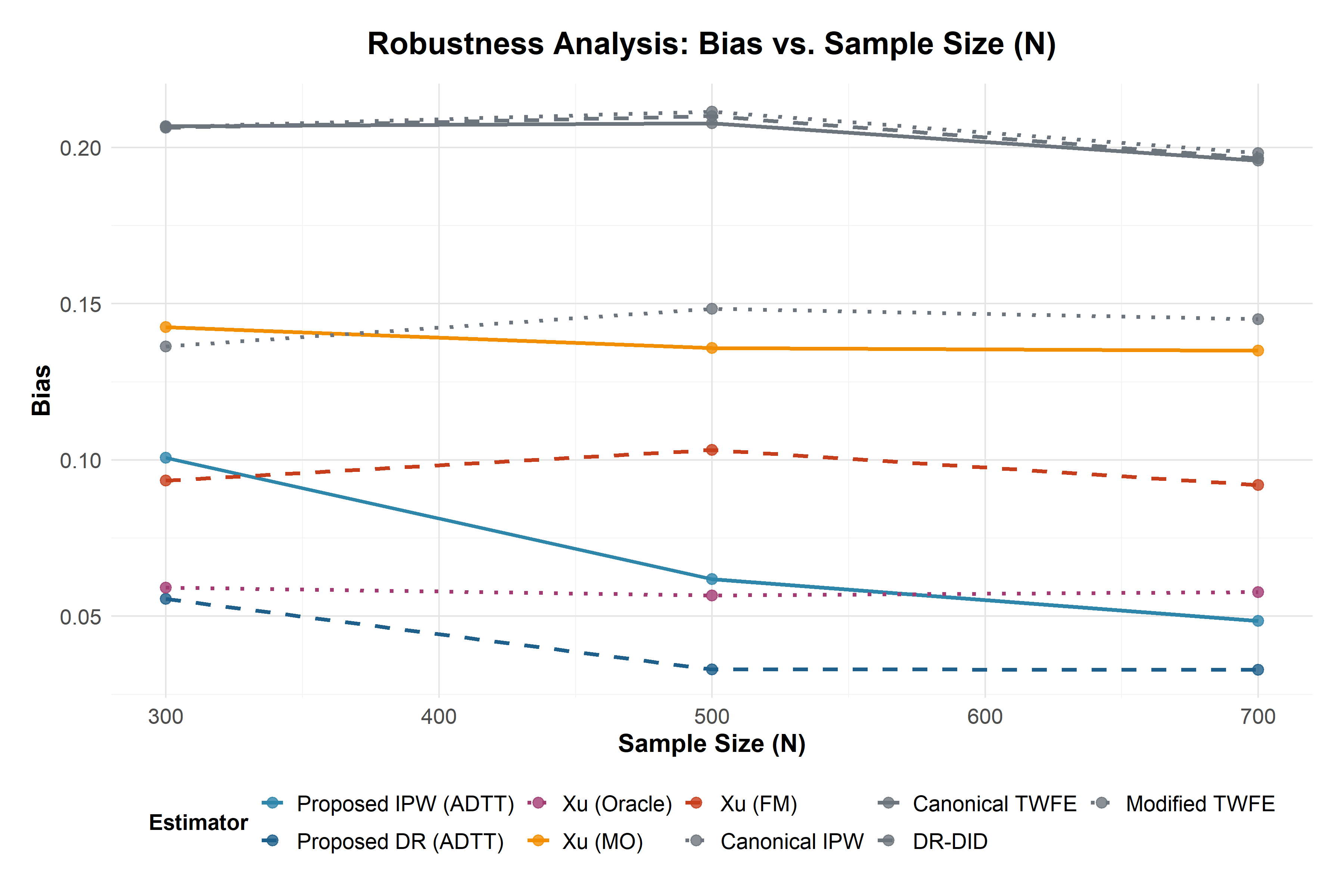}
\caption{Bias of the proposed IPW and DR ADTT estimators and benchmark methods as a function of sample size $N \in \{300, 500, 700\}$. The proposed IPW and DR methods show decreasing bias with increasing sample size, providing empirical evidence for consistency. Benchmark methods that ignore interference or rely on misspecified exposure mappings retain substantial bias even with larger sample sizes.}
\label{fig:robustness_sample_size_bias}
\end{figure}

We also examine the asymptotic properties of the estimators by varying the sample size $n \in \{300, 500, 700\}$. 
Figure~\ref{fig:robustness_sample_size_bias} shows the Bias as a function of $n$.
As expected, the performance of the proposed IPW and DR ADTT estimators improves as the sample size increases, with both Bias and RMSE decreasing towards zero. This provides empirical evidence for the consistency of our estimators (Theorem \ref{thm:consistency_ADTT}). In contrast, the estimators that ignore interference or rely on misspecified exposure mappings retain substantial bias even with larger sample sizes, indicating their inconsistency under this DGP.

\subsection{Robustness checks}

We conduct additional robustness checks to examine how our proposed method performs under different network structures and dependence patterns. These experiments provide further evidence of the robustness of our nonparametric approach.

\begin{figure}[!htbp]
\includegraphics[width=1.0\linewidth]{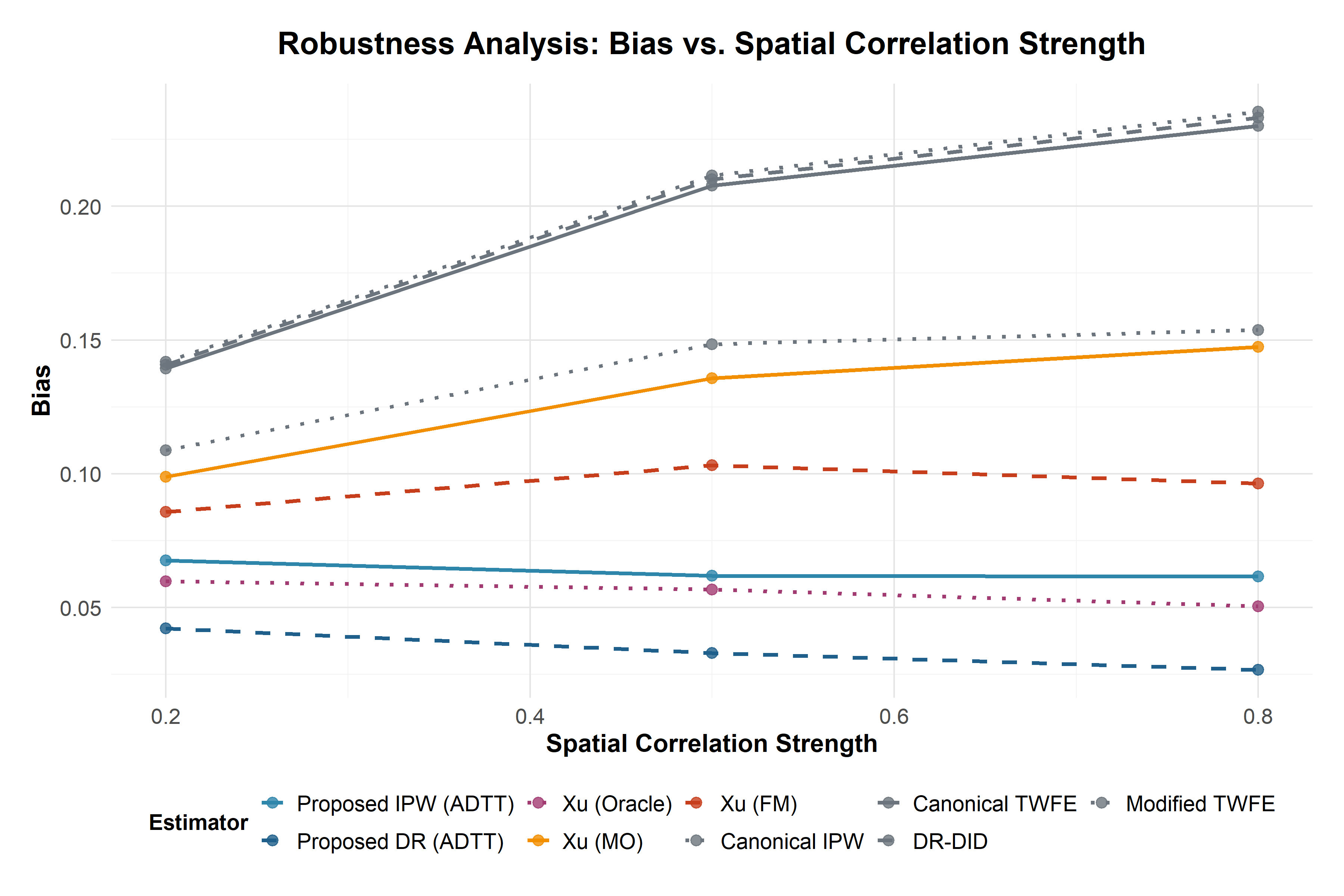}
\caption{Bias as a function of spatial correlation strength ($\rho_0 \in \{0.2, 0.5, 0.8\}$).}
\label{fig:robustness_correlation_bias}
\end{figure}

To check the effect of network dependence, we vary the strength of spatial correlation in the unobserved confounder by changing the base correlation parameter $\rho_0 \in \{0.2, 0.5, 0.8\}$. Stronger spatial correlation increases the dependence between neighboring units, making it more challenging to identify causal effects. Figure~\ref{fig:robustness_correlation_bias} demonstrates that our proposed method remains robust across different levels of spatial correlation, while benchmark methods that ignore interference show deteriorating performance as correlation increases.

\begin{figure}[!htbp]
\includegraphics[width=1.0\linewidth]{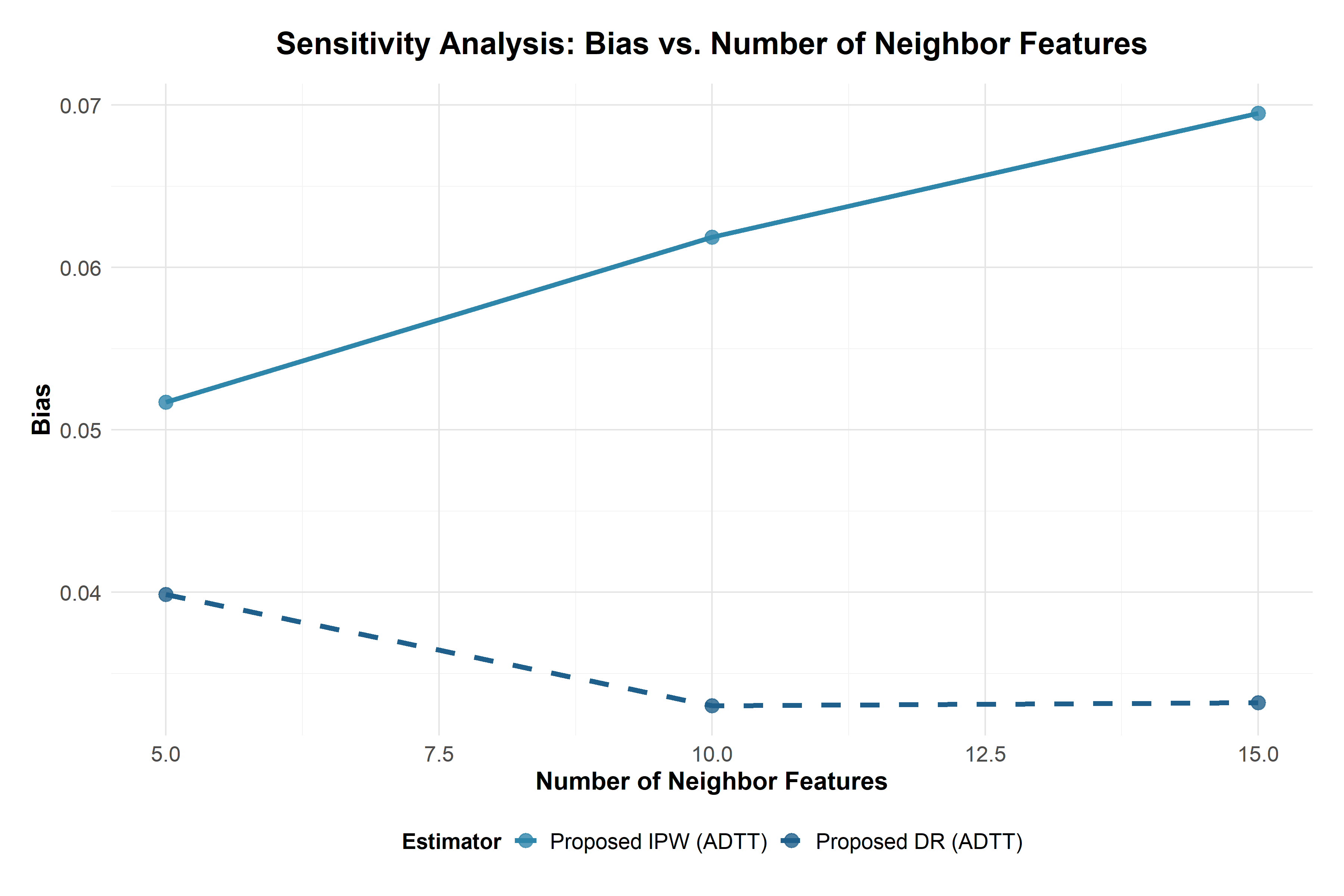}
\caption{Bias as a function of the number of neighborhood features used in propensity score estimation.}
\label{fig:sensitivity_features_bias}
\end{figure}

Next, we investigate sensitivity of neighborhood size of the proposed method.
A potential concern with our approach is the curse of dimensionality when neighborhood sizes are large. To examine this, we conduct a sensitivity analysis where we vary the number of neighborhood features used in the propensity score estimation. Figure \ref{fig:sensitivity_features_bias} shows that while the IPW estimator's performance degrades slightly as the effective dimensionality increases, the DR estimator maintains stable or even improved performance, demonstrating the robustness of the doubly robust approach against the curse of dimensionality. Both methods maintain reasonable performance even with larger neighborhoods, demonstrating the practical feasibility of our approach.

\section{Real Data Application: SEZ Policy in China}

We apply our method to evaluate the effects of China's Special Economic Zone (SEZ) policy, a prominent place-based policy intervention. This application demonstrates the practical importance of accounting for spillover effects in policy evaluation and illustrates how our identification-based approach can be applied in real-world settings where the interference structure is unknown.

We use village-level data from \citet{Xu2025} for China (2004 pre-treatment, 2008 post-treatment). The treatment variable $D_i$ indicates whether village $i$ was designated as an SEZ.
Since geographic coordinates are unavailable, we define neighborhoods at the county level using covariate similarity. We use the same model specification as in Section 3.1 with $L=10$. For benchmarks, we construct an exposure mapping based on the leave-one-out SEZ ratio within each county. We define exposure as the number of treated neighbors (SEZ villages) within each village's neighborhood. Villages are classified as ``Low Exposure" if their exposure is below the median, and ``High Exposure" if at or above the median.
We analyze three outcome variables in log scale, capital, employment, and output. 
We implement the proposed IPW estimators as described in Section 3.1, using covariate similarity to define neighbor ranks. We compare our proposed estimators with methods from \citet{Xu2025} (requiring exposure mapping) and standard DID estimators that ignore interference. Standard errors are computed using a HAC estimator that accounts for within-county spatial correlation based on covariate similarity. We use covariate similarity as a practical adaptation since geographic distance is unavailable within counties.

Table~\ref{tab:sez_results} presents the results. SEZ policy shows positive direct effects, with larger impacts on villages with low exposure. We find negative spillover effects among treated villages, suggesting competition effects. Canonical DID estimates are substantially larger than estimates accounting for spillovers, indicating positive selection bias. The difference between the IPW and DR estimates for ADTT (0.2318 vs 0.0283) may reflect differences in model specification or finite-sample properties; both estimators target the same ADTT estimand under their respective identifying assumptions.

\begin{table}[h]
\centering
\caption{Estimated direct and spillover effects of SEZ Policy. Standard errors in parentheses (HAC estimator). All outcomes in log scale.
}
\label{tab:sez_results}
\begin{tabular}{lccc}
\hline
& \multicolumn{3}{c}{Outcome Variable} \\
\cline{2-4}
Effect & Capital (log) & Employment (log) & Output (log) \\
\hline
\multicolumn{4}{l}{\textit{Direct Effects by Exposure Level}} \\
ADTT (Low Exposure) & 0.0356 & 0.0309 & 0.0362 \\
ADTT (High Exposure) & 0.0237 & 0.0197 & 0.0245 \\
\hline
\multicolumn{4}{l}{\textit{Average Direct Effect (ADTT)}} \\
ADTT (DR) & 0.0283 & 0.0240 & 0.0290 \\
\textbf{Proposed IPW (ADTT)} & \textbf{0.2318} & \textbf{0.2629} & \textbf{0.2066} \\
& \textbf{(0.0269)} & \textbf{(0.0181)} & \textbf{(0.0300)} \\
Canonical DID & 0.3552 & 0.3464 & 0.3582 \\
\hline
\multicolumn{4}{l}{\textit{Indirect Effects}} \\
\textbf{Proposed IPW (AITT)} & \textbf{-0.2174} & \textbf{-0.1015} & \textbf{-0.2445} \\
& \textbf{(0.0165)} & \textbf{(0.0131)} & \textbf{(0.0209)} \\
\hline
\multicolumn{4}{l}{\textit{Spillover Effects}} \\
Spillover (Treated) & -0.1930 & -0.1438 & -0.1955 \\
Spillover (Control) & -0.0027 & 0.0328 & -0.0023 \\
\hline
\end{tabular}
\smallskip
\footnotesize
\end{table}

\section{Concluding Remarks}

We proposed a DID framework under local network interference that does not rely on a known exposure mapping. Identification is nonparametric under conditional parallel trends given the neighborhood treatment vector; the AITT estimand captures the externality of the intervention on neighbors. Asymptotic theory for the IPW estimators rests on weakly dependent random fields, and we gave a consistent variance estimator for inference.
Our simulation studies demonstrated the robustness of our identification-based approach (nonparametric in identification) in settings with complex and unknown interference structures, outperforming methods that rely on misspecified exposure mappings or ignore interference altogether. The real data application to China's SEZ policy further illustrated the practical importance of accounting for spillover effects, revealing substantial policy effects and important heterogeneity patterns that would be missed by standard methods.

We stress that further study is needed to obtain a clearer picture of when our approach is preferred over exposure mapping methods in finite samples, particularly as network density or neighborhood size varies. Extending this framework to settings with staggered treatment adoption would combine our setting with the challenges recently highlighted in the DID literature; while such an extension is of clear interest, it makes identification and inference more involved, so we leave it for future work. Another important direction is the development of methods for estimating propensity scores nonparametrically when the neighborhood size is large, potentially leveraging machine learning techniques to address the curse of dimensionality.

\section*{Acknowledgement}
This work is supported by the Japan Society for the Promotion of Science (JSPS KAKENHI) grant numbers, 24K00244 and 25H00546. 

\vspace{0.5cm}
\bibliographystyle{chicago}
\bibliography{ref.bib}

\newpage
\setcounter{equation}{0}
\setcounter{section}{0}
\setcounter{table}{0}
\setcounter{figure}{0}
\setcounter{page}{1}
\renewcommand{\thesection}{S\arabic{section}}
\renewcommand{\theequation}{S\arabic{equation}}
\renewcommand{\thetable}{S\arabic{table}}
\renewcommand{\thefigure}{S\arabic{figure}}

\vspace{1cm}
\begin{center}
{\LARGE
{\bf Supplementary Material for ``Difference-in-Differences under Local Dependence on Networks"}
}
\end{center}

\medskip
This Supplementary Material provides proofs of the theorems in the main text. \

\section{Proof of Theorem \ref{thm:identification_ADTT}}

By the definition of ADTT, we marginalize over all possible combinations of the neighborhood treatment vector $\boldsymbol{d} = \boldsymbol{d}_{N_i}$ for each unit $i$.
\begin{align*}
\tau_{\mathrm{dir}} &= \frac{1}{n} \sum_{i=1}^n E[ y_{2i} (1, \bD_{N_i}) - y_{2i} (0, \bD_{N_i}) | D_i = 1, z_i] \\
&= \frac{1}{n} \sum_{i=1}^n \sum_{\boldsymbol{d}} E[ y_{2i} (1, \boldsymbol{d}) - y_{2i} (0, \boldsymbol{d}) | D_i = 1, \bD_{N_i} = \boldsymbol{d}, z_i] \\
& \times P(\bD_{N_i} = \boldsymbol{d} | D_i = 1, z_i)
\end{align*}
The conditional expectation term can be rewritten as the difference in trends, 
\begin{align*}
& E[ y_{2i} (1, \boldsymbol{d}) - y_{2i} (0, \boldsymbol{d}) | D_i = 1, \bD_{N_i} = \boldsymbol{d}, z_i] \\
& = E[ y_{2i} (1, \boldsymbol{d}) - y_{1i} | D_i = 1, \bD_{N_i} = \boldsymbol{d}, z_i] \\
& - E[ y_{2i} (0, \boldsymbol{d}) - y_{1i} | D_i = 1, \bD_{N_i} = \boldsymbol{d}, z_i]
\end{align*}
By Assumption \ref{as:trend}, the second term (the counterfactual trend) is identified by the observed trend in the control group ($D_i=0$):
\begin{align*}
E[ y_{2i} (0, \boldsymbol{d}) - y_{1i} | D_i = 1, \bD_{N_i} = \boldsymbol{d}, z_i] = E[ Y_{2i} - Y_{1i} | D_i = 0, \bD_{N_i} = \boldsymbol{d}, z_i]
\end{align*}
Let $\Delta_{i,d} = E[ Y_{2i} - Y_{1i} | D_i=1, \bD_{N_i}=\boldsymbol{d}, z_i] - E[ Y_{2i} - Y_{1i} | D_i=0, \bD_{N_i}=\boldsymbol{d}, z_i]$.
Also, by Bayes' theorem, 
\begin{align*}
P(\bD_{N_i} = \boldsymbol{d} | D_i = 1, z_i) 
& =  \frac{P(D_i=1|\bD_{N_i}=\boldsymbol{d}, z_i)P(\bD_{N_i}=\boldsymbol{d}|z_i)}{P(D_i=1|z_i)} \\
& = \frac{e_i(\boldsymbol{d}, z_i)P(\bD_{N_i}=\boldsymbol{d}|z_i)}{\pi_i(z_i)}   
\end{align*}

Thus, ADTT can be rewritten using an IPW-like form. Let $e_i = e_i(\bD_{N_i}, z_i)$.
\begin{align*}
\tau_{\mathrm{dir}} &= \frac{1}{n} \sum_{i=1}^n \sum_{\boldsymbol{d}} \Delta_{i,d} \frac{e_i(\boldsymbol{d}, z_i)P(\bD_{N_i}=\boldsymbol{d}|z_i)}{\pi_i(z_i)} \\
&= \frac{1}{n} \sum_{i=1}^n E \left[ \left( E \left[ \left( \frac{D_i}{e_i} - \frac{1-D_i}{1-e_i} \right) (Y_{2i}-Y_{1i})|\bD_{N_i}, z_i\right] \right) \frac{e_i}{\pi_i} \middle| z_i \right] \\
&= \frac{1}{n} \sum_{i=1}^n E \left[ E \left[ \frac{D_i-e_i}{e_i(1-e_i)}(Y_{2i}-Y_{1i}) \middle| \bD_{N_i}, z_i \right] \frac{e_i}{\pi_i} \middle| z_i \right] \\
&= \frac{1}{n} \sum_{i=1}^n E \left[ \frac{D_i-e_i}{\pi_i(1-e_i)}(Y_{2i}-Y_{1i}) \middle| z_i \right]
\end{align*}
This completes the proof.

\section{Proof of Theorem \ref{thm:identification_AITT}}
We consider the expectation term in the definition of AITT for a specific neighbor $j \in N_i$. Let $\bD^{d} = \bD_{N_j}^{(d)}$, $\bD^{-i} = \bD_{N_j}^{-i}$. We aim to identify:
$$
\tau_{ij} = E[ y_{2j} (D_j, \bD^{(1)}) - y_{2j} (D_j, \bD^{(0)}) | D_i = 1, z_i, z_j],
$$
where $\bD^{(d)}$ denotes the treatment vector of the neighbors of $j$ with $D_i=d\in \{0,1\}$.
We marginalize over $D_j$ and $\bD^{-i}$.
\begin{align*}
\tau_{ij} = \sum_{d_j, \boldsymbol{d}^{-i}} & E[ y_{2j} (d_j, D_i=1, \boldsymbol{d}^{-i}) - y_{2j} (d_j, D_i=0, \boldsymbol{d}^{-i}) | D_i = 1, D_j=d_j, \bD_{N_j}^{-i} = \boldsymbol{d}^{-i}, z_i, z_j] \\
& \times P(D_j = d_j, \bD_{N_j}^{-i} = \boldsymbol{d}^{-i} | D_i = 1, z_i, z_j).
\end{align*}

Let $\text{CATT}_{ij}(d_j, \boldsymbol{d}^{-i})$ denote the conditional expectation term. We rewrite it as the difference in trends:

\begin{align*}
\text{CATT}_{ij}(d_j, \boldsymbol{d}^{-i}) 
& = E[ y_{2j} (d_j, D_i=1, \boldsymbol{d}^{-i}) - y_{1j} | D_i = 1, D_j=d_j, \bD_{N_j}^{-i} = \boldsymbol{d}^{-i}, z_i, z_j]\\
& - E[ y_{2j} (d_j, D_i=0, \boldsymbol{d}^{-i}) - y_{1j} | D_i = 1, D_j=d_j, \bD_{N_j}^{-i} = \boldsymbol{d}^{-i}, z_i, z_j]
\end{align*}

We apply Assumption \ref{as:trend} to unit $j$ to identify the counterfactual trend:
\begin{align*}
& E[ y_{2j} (d_j, D_i=0, \boldsymbol{d}^{-i}) - y_{1j} | D_i = 1, D_j=d_j, \bD_{N_j}^{-i} = \boldsymbol{d}^{-i}, z_i, z_j]\\
& = E[ y_{2j} (d_j, D_i=0, \boldsymbol{d}^{-i}) - y_{1j} | D_i = 0, D_j=d_j, \bD_{N_j}^{-i} = \boldsymbol{d}^{-i}, z_i, z_j] \\
&= E[ Y_{2j} - Y_{1j} | D_i = 0, D_j=d_j, \bD_{N_j}^{-i} = \boldsymbol{d}^{-i}, z_i, z_j].
\end{align*}

Thus, $\text{CATT}_{ij}(d_j, \boldsymbol{d}^{-i})$ is identified by the conditional DID for unit $j$'s outcome with respect to unit $i$'s treatment. Similar to Theorem \ref{thm:identification_ADTT}, this is expressed in IPW form using 

$$e'_{ij}(d_j, \boldsymbol{d}^{-i}, z_i, z_j) = P(D_i=1 | D_j=d_j, \bD_{N_j}^{-i}=\boldsymbol{d}^{-i}, z_i, z_j)$$. 

Letting $e'_{ij} = e'_{ij}(d_j, \boldsymbol{d}^{-i}, z_i, z_j)$, we have
$$
\text{CATT}_{ij}(d_j, \boldsymbol{d}^{-i}) = E \left[ \frac{D_i - e'_{ij}}{e'_{ij}(1-e'_{ij})} ( Y_{2j} - Y_{1j} ) \middle| D_j=d_j, \bD_{N_j}^{-i}=\boldsymbol{d}^{-i}, z_i, z_j \right].
$$
Now we substitute this back into $\tau_{ij}$. We use Bayes' theorem:
$$
P(D_j = d_j, \bD_{N_j}^{-i} = \boldsymbol{d}^{-i} | D_i = 1, z_i, z_j) = \frac{e'_{ij} P(D_j = d_j, \bD_{N_j}^{-i} = \boldsymbol{d}^{-i} | z_i, z_j)}{P(D_i=1|z_i)}.
$$
\begin{align*}
\tau_{ij} 
&= E \left[ E \left[ \frac{D_i - e'_{ij}}{e'_{ij}(1-e'_{ij})} ( Y_{2j} - Y_{1j} ) \middle| D_j, \bD_{N_j}^{-i}, z_i, z_j \right] \times \frac{e'_{ij}}{\pi_i(z_i)} \middle| z_i, z_j \right] \\
&= E \left[ \frac{D_i - e'_{ij}}{\pi_i(z_i)(1-e'_{ij})} ( Y_{2j} - Y_{1j} ) \middle| z_i, z_j \right].
\end{align*}
Finally, averaging over $j \in N_i$ and $i \in I_n$ yields the identification result for AITT.

\section{Proof of Lemma \ref{lem:psi_ADTT}}

\subsection{Proof for for ADTT}
Let $\phi_i = \phi_i^{\mathrm{dir}} = \frac{D_i - e_i(z_i)}{\pi_i(z_i) (1-e_i(z_i))} ( Y_{2i} - Y_{1i} )$. We verify the definition of $\psi$-dependence.
Consider any $h, h^{\prime} \in \mathbb{N}$, bounded Lipschitz functions $f \in \mathcal{L}_h, f^{\prime} \in \mathcal{L}_{h^{\prime}}$, and any pair of sets $(H, H^{\prime}) \in \mathcal{P}_n(h, h^{\prime} ; s)$ such that $l_A(H, H') \geq s > 0$.

\noindent
Case 1: $s \leq 2K$. By Assumptions \ref{as:overlap} (Overlap) and \ref{as:bounded_outcomes} (Bounded Outcomes), $\{\phi_i\}$ are uniformly bounded. Since $f$ and $f'$ are bounded, the covariance is bounded:
$$
|\operatorname{Cov}(f(\phi_H), f^\prime (\phi_{H^\prime}))| \leq 4 \|f\|_{\infty} \|f^{\prime}\|_{\infty}.
$$

\noindent
Case 2: $s > 2K$. Let $x_i = (Y_{1i}, D_i, \epsilon_i)$ be the fundamental variables. By Assumption \ref{as:local_interference} (Local Interference), $\phi_i$ is a measurable function of the variables in the $K$-neighborhood of $i$:
$$
\phi_i = g(\{ x_k \}_{k \in N(i; K)^+}).
$$
By Assumptions \ref{as:overlap} and \ref{as:bounded_outcomes}, $g$ is Lipschitz continuous.

Let $H^K = \bigcup_{i \in H} N(i; K)^+$. Then, $\sigma(\{ \phi_i \}_{i \in H}) \subset \sigma(\{ x_i \}_{i \in H^K})$.
When $s > 2K$, the distance between $H^K$ and $(H')^K$ is at least $s-2K$.
Since $\{x_i\}$ is $\alpha$-mixing (Assumption \ref{as:alpha_mixing}), we apply Proposition 2.2 in \citet{Kojevnikov2021} to bound the covariance:
\begin{align*}
|\operatorname{Cov}(f(\phi_H), f^\prime (\phi_{H^{\prime}}))| &\leq 4 \|f\|_{\infty} \|f^{\prime}\|_{\infty} \alpha(\sigma(\{x_i\}_{i \in H^K}), \sigma(\{x_i\}_{i \in (H')^K})) \\
&\leq 4 \|f\|_{\infty} \|f^{\prime}\|_{\infty} \bar{\alpha} (|H^K|, |(H')^K|, s - 2K).
\end{align*}
Using Assumption \ref{as:alpha_mixing}, $\bar{\alpha}(h, h', s) \leq \psi(h, h') \widehat{\alpha}(s)$. Assuming the neighborhood size $|N(i;K)^+|$ is bounded by $C_K$, we have $|H^K| \leq h C_K$.
\begin{align*}
|\operatorname{Cov}(f(\phi_H), f^\prime (\phi_{H^{\prime}}))| &\leq 4 \|f\|_{\infty} \|f^{\prime}\|_{\infty} \psi(h C_K, h' C_K) \widehat{\alpha}(s-2K).
\end{align*}
Combining both cases, we define the coefficients. Let $\Psi(h, h') = \max \{1, \psi(h C_K, h' C_K) \}$. We set:
$$
\psi_{h, h^\prime} (f, f^\prime) = 4 \|f\|_{\infty} \|f^{\prime}\|_{\infty} \Psi(h, h'),
$$
$$
\tilde{\theta}_{n, s} = \mathbf{1} \{ s \leq 2K \} + \mathbf{1} \{ s > 2K \} \widehat{\alpha} (s-2K).
$$
This satisfies the definition of $\psi$-dependence.

\subsection{Proof for AITT}
Define
$$
\phi_i = \phi_i^{\mathrm{ind}} = \frac{1}{|N_i|} \sum_{j \in N_i} \left\{ \frac{D_i - e'_{ij}(z_i, z_j)}{\pi_i(z_i)(1-e'_{ij}(z_i, z_j))} (Y_{2j} - Y_{1j}) \right\},
$$
and $x_i = (Y_{1i}, D_i, \epsilon_i)$. The term $(Y_{2j}-Y_{1j})$ and $e'_{ij}$ depend on $\{x_k\}_{k \in N_j^+}$, where $N_j^+ = N_j \cup \{j\}$.
By Assumption \ref{as:neighborhood_sufficiency}, $j \in N_i$ implies $l_A(i,j) \leq K$, and similarly for $k \in N_j^+$, we have $l_A(j,k) \leq K$ (since $N_j$ consists of the $L$ nearest neighbors of $j$). Therefore, the distance between $i$ and any $k \in N_j^+$ is at most $l_A(i,j)+l_A(j,k) \leq K+K = 2K$.
Therefore, $\phi_i$ is a function of the variables in the $2K$-neighborhood of $i$, defined as $\phi_i = g(\{ x_k \}_{k \in N(i; 2K)^+})$.
By Assumptions \ref{as:overlap} and \ref{as:bounded_outcomes}, $g$ is Lipschitz continuous and bounded.

Consider sets $H, H'$ such that $l_A(H, H') \geq s$.
If $s \leq 4K$, the covariance is bounded by $4 \|f\|_{\infty} \|f^{\prime}\|_{\infty}$.
If $s > 4K$, let $H^{2K} = \bigcup_{i \in H} N(i; 2K)^+$. The distance between $H^{2K}$ and $(H')^{2K}$ is at least $s-4K$.
Applying Proposition 2.2 in \citet{Kojevnikov2021} and Assumption \ref{as:alpha_mixing}, we have 
\begin{align*}
|\operatorname{Cov}(f(\phi_H), f^\prime (\phi_{H^{\prime}}))| &\leq 4 \|f\|_{\infty} \|f^{\prime}\|_{\infty} \bar{\alpha} (|H^{2K}|, |(H')^{2K}|, s - 4K) \\
&\leq 4 \|f\|_{\infty} \|f^{\prime}\|_{\infty} \psi(|H^{2K}|, |(H')^{2K}|) \widehat{\alpha}(s-4K).
\end{align*}
Assuming bounded neighborhood sizes, we can define the $\psi$-dependence coefficients similarly to Lemma \ref{lem:psi_ADTT}, with the decay rate, as follows: 
$$
\tilde{\theta}_{n,s} = \mathbf{1} \{ s \leq 4K \} + \mathbf{1} \{ s > 4K \} \widehat{\alpha} (s-4K),
$$
which completes the proof.

\section{Proofs of Main Theorems}

\subsection{Proof of Theorem~\ref{thm:consistency_ADTT}}
The result follows directly from Theorem 3.1 in \citet{Kojevnikov2021}, given that the summand is $\psi$-dependent (Lemma \ref{lem:psi_ADTT}) and satisfies the weak dependence condition for the LLN (Assumption \ref{as:LLN}).

\subsection{Proof of Theorem \ref{thm:normality_ADTT}}
From Lemma \ref{lem:psi_ADTT} and Assumption \ref{as:CLT}, the conditions for Theorem 3.2 (CLT) in \citet{Kojevnikov2021} are satisfied, which establishes asymptotic normality.

\subsection{Proof of Theorem \ref{thm:HAC}}
Since Lemmas \ref{lem:psi_ADTT} show that $\phi_i$ is $\psi$-dependent, the consistency is established by applying Proposition 4.1 of \citet{Kojevnikov2021} under Assumption \ref{as:HAC}.

\subsection{Proof of Theorem \ref{thm:DR_robustness}}
We prove the double robustness property only for the ADTT estimator since the proof for AITT can be done in the same way. 
The DR estimator for ADTT can be written as:
\begin{align*}
\widehat{\tau}_{\mathrm{dir}}^{DR} 
&= \frac{1}{n} \sum_{i=1}^n \Bigg\{ \frac{D_i}{\widehat{\pi}_i(z_i)}\left(\Delta Y_i-\Delta m_{1i}(\bD_{N_i}, z_i)\right) 
-\frac{(1-D_i)\widehat{e}_i(z_i)}{\widehat{\pi}_i(z_i)(1-\widehat{e}_i(z_i))}\left(\Delta Y_i-\Delta m_{0i}(\bD_{N_i}, z_i)\right) \\
& \quad  \quad 
+ \frac{\widehat{e}_i(z_i)}{\widehat{\pi}_i(z_i)}\left(\Delta m_{1i}(\bD_{N_i}, z_i)-\Delta m_{0i}(\bD_{N_i}, z_i) \right)\Bigg\}.
\end{align*}

\noindent
\textbf{Case 1: Propensity score models are correctly specified.}
When $e_i = \widehat{e}_i$ and $\pi_i = \widehat{\pi}_i$ are correctly specified, we can show that the regression terms cancel out. Specifically, by the law of iterated expectations and the fact that $E[D_i | \bD_{N_i}, z_i] = e_i(\bD_{N_i}, z_i)$, we have:
\begin{align*}
E\left[ \frac{D_i}{\widehat{\pi}_i(z_i)} \Delta m_{1i}(\bD_{N_i}, z_i) \right] &= E\left[ \frac{1}{\pi_i(z_i)} E[D_i \Delta m_{1i}(\bD_{N_i}, z_i) | \bD_{N_i}, z_i] \right] \\
&= E\left[ \frac{e_i(\bD_{N_i}, z_i)}{\pi_i(z_i)} \Delta m_{1i}(\bD_{N_i}, z_i) \right] \\
&= E\left[ \frac{D_i}{\pi_i(z_i)} (Y_{2i} - Y_{1i}) \right],
\end{align*}
and similarly for the $\Delta m_{0i}$ term. 
The third term $\widehat{e}_i(z_i)\left(\Delta m_{1i}-\Delta m_{0i}\right)/\widehat{\pi}_i(z_i)$ has expectation
\begin{align*}
&E\left[ \frac{\widehat{e}_i(z_i)}{\widehat{\pi}_i(z_i)}\left(\Delta m_{1i}(\bD_{N_i}, z_i)-\Delta m_{0i}(\bD_{N_i}, z_i)\right) \right] \\
& \quad \quad 
= E\left[ \frac{e_i(\bD_{N_i}, z_i)}{\pi_i(z_i)} \left(\Delta m_{1i}(\bD_{N_i}, z_i)-\Delta m_{0i}(\bD_{N_i}, z_i)\right) \right],
\end{align*}
which results in cancellation of the model prediction terms when combined with the first two terms. 
The DR estimator then reduces to the IPW estimator, which is consistent under Assumptions \ref{as:NoAnticipation} and \ref{as:trend} by Theorem \ref{thm:identification_ADTT}.

\noindent
\textbf{Case 2: Outcome regression models are correctly specified.}
When $\mu_{1i} = \Delta m_{1i}$ and $\mu_{0i} = \Delta m_{0i}$ are correctly specified, we can show that the IPW residual terms have zero expectation. Specifically, by the law of iterated expectations:
\begin{align*}
&E\left[ \frac{D_i}{\pi_i(z_i)}\left(\Delta Y_i-\Delta m_{1i}(\bD_{N_i}, z_i)\right) \middle| \bD_{N_i}, z_i \right]\\
& \quad \quad
= \frac{e_i(\bD_{N_i}, z_i)}{\pi_i(z_i)} \left(E[\Delta Y_i | D_i=1, \bD_{N_i}, z_i] - \mu_{1i}(\bD_{N_i}, z_i)\right) = 0,
\end{align*}
and similarly for the second term. 
The third term $\widehat{e}_i(z_i)\left(\Delta m_{1i}-\Delta m_{0i}\right)/\widehat{\pi}_i(z_i)$ has expectation
\begin{align*}
&E\left[ \frac{\widehat{e}_i(z_i)}{\widehat{\pi}_i(z_i)}\left(\Delta m_{1i}(\bD_{N_i}, z_i)-\Delta m_{0i}(\bD_{N_i}, z_i)\right) \right] \\
& \quad = E\left[ \frac{\widehat{e}_i(z_i)}{\widehat{\pi}_i(z_i)} \left(\mu_{1i}(\bD_{N_i}, z_i)-\mu_{0i}(\bD_{N_i}, z_i)\right) \right].
\end{align*}
When the outcome regression models are correctly specified, this expectation equals ADTT, as the expectation is taken over the distribution of $\bD_{N_i}$ conditional on $D_i=1$ (captured by the weight $\widehat{e}_i(z_i)/\widehat{\pi}_i(z_i)$). The DR estimator then reduces to this regression-based estimator, which is consistent under Assumptions \ref{as:NoAnticipation} and \ref{as:trend}.
This establishes the double robustness property that the DR estimator is consistent if either the propensity score models or the outcome regression models are correctly specified.

\subsection{Proof of Theorem \ref{thm:DR_normality}}
When both the propensity score models and outcome regression models are correctly specified, the DR estimator can be shown to be asymptotically normal using similar arguments to Theorem \ref{thm:normality_ADTT}. The key is to establish that the DR influence function $\phi_i^{DR}$ defined in equation (\ref{eq:Z_DR}) satisfies the $\psi$-dependence property (similar to Lemma \ref{lem:psi_ADTT}) and the weak dependence conditions (Assumptions \ref{as:CLT}). 

Since $\phi_i^{DR}$ is a function of the same fundamental variables $(Y_{1i}, D_i, \epsilon_i)$ and their neighborhoods, and the outcome regression functions $\mu_{1i}$, $\mu_{0i}$, etc. are bounded under Assumption \ref{as:bounded_outcomes}, the $\psi$-dependence structure follows from the same arguments as in Lemma \ref{lem:psi_ADTT}. The asymptotic normality then follows from Theorem 3.2 in \citet{Kojevnikov2021} under Assumption \ref{as:CLT}.

\end{document}